\documentclass[10pt]{article}

\usepackage{amsmath}
\usepackage{amsfonts}
\usepackage{amssymb}

\usepackage{graphicx}
\usepackage{caption}
\usepackage{subcaption}
\usepackage{lineno}
\usepackage{float}

\usepackage{verbatim}

\usepackage{color}
\usepackage{hyperref}
\usepackage{stackrel}

\usepackage{upgreek}
\usepackage{csquotes}
\usepackage{ulem}
 \usepackage{cite}

 \def\##1{{\bf #1}}

\def\=#1{\underline{\underline{#1}}}

\def\quote#1{\textquotedblleft #1\textquotedblright}

\def\red{\textcolor{red}}

\definecolor{Lak}{rgb}{0.7,0.3,0.7}

 \def\eps{\varepsilon}

\def\epso{\eps_{\scriptscriptstyle 0}}

\def\epsr{\eps_{\rm r}}

\def\ux{\hat{\#x}}
\def\uy{\hat{\#y}}
\def\uz{\hat{\#z}}

\def\up{\hat{\#p}}

\def\fscaone{f_{\rm pert}^{(1)}}

\def\fscatwo{f_{\rm pert}^{(2)}}

\def\bro{{\#r}_{\rm o}}
\def\ro{r_{\rm o}}
\def\thetao{\theta_{\rm o}}
\def\phio{\phi_{\rm o}}
\def\thetap{\theta_{\rm p}}
\def\phip{\phi_{\rm p}}

\def\.{\mbox{ \tiny{$^\bullet$} }}

\def\les{\left[}
\def\ris{\right]}
\def\lec{\left\{}
\def\ric{\right\}}

\def\calA{{\cal A}}
\def\calB{{\cal B}}
\def\calC{{\cal C}}

\def\calT{{\cal T}}
\def\calQ{{\cal Q}}

\def\alphax{\alpha_{\rm x}}
\def\alphay{\alpha_{\rm y}}

\def\alphas{\alpha_{\rm s}}
\def\betas{\beta_{\rm s}}
\def\gammas{\gamma_{\rm s}}

\def\aave{a_{\rm ave}}
\def\epsave{\eps_{\rm ave}}

\def\rtilde{\tilde{r}}

\def\inc{_{\rm so}}
\def\sca{_{\rm pert}}
\def\ind{_{\rm int}}

\def\smn{_{\rm smn}}
\def\smnp{_{\rm s^\prime m^\prime n^\prime}}
\def\double{_{\rm ss^\prime mm^\prime nn^\prime}}

\def\emn{_{\rm emn}}
\def\omn{_{\rm omn}}

\def\Emn{E_{\rm mn}}

\def\out{_{\rm out}}
\def\ins{_{\rm in}}
\def\Vout{{\sf V}\out}
\def\Vin{{\sf V}\ins}
\def\Sinf{{\sf S}_\infty}
\def\So{{\sf S}}
\def\Vso{{\sf V}_{\rm so}}

\def\rout{r\out}
\def\rin{r\ins}
\def\aout{a\out}
\def\ain{a\ins}
\def\rso{r_{\rm so}}
\def\rinf{r_{\infty}}

\def\nhat{\hat{\#n}}
\def\nr{\nhat(\#r)}
\def\nrp{\nhat(\#r^\prime)}

\def\plus{_{+}}
\def\minus{_{-}}

\def\epsrel{\=\eps_{\,{\rm rel}}}

\begin{document}

\begin{center}
\textbf{Theory of perturbation of electric potential by a 3D object made of an anisotropic dielectric material}\\
\vspace{10pt}
{Akhlesh Lakhtakia,$^{1}$ Hamad M. Alkhoori$^{2}$ and Nikolaos L. Tsitsas$^3$}\\
\vspace{10pt}

{$^1$Department of Engineering Science and Mechanics, 
The Pennsylvania State University, University Park, Pennsylvania 16802, USA}\\
{$^2$Department of Electrical Engineering, United Arab Emirates University, P.O. Box 15551, Al Ain, UAE}\\
{$^3$School of Informatics, Aristotle University of Thessaloniki, Thessaloniki 54124, Greece}

\vspace{10pt}

\end{center}

\begin{abstract}
The extended boundary condition method (EBCM) was formulated for the
perturbation of a source electric potential by a
 3D object  composed of a homogeneous anisotropic dielectric medium whose relative permittivity
dyadic is positive definite. The formulation required the application of Green's second identity to
the exterior region to deduce the electrostatic counterpart of the Ewald--Oseen extinction theorem.
The electric potential inside the object was represented using a basis obtained by implementing
an affine bijective transformation of space to the Gauss equation for the electric field. The EBCM
yields a transition matrix that depends on the geometry and the composition of the 3D object,
but not on the source potential.

\end{abstract}

\vspace{2pc}
\vspace{2pc}

\section{Introduction} Dating back more than a hundred years,
the Ewald--Oseen extinction theorem \cite{Ewald1916,Oseen1915} states that when a 3D object is illuminated by an
incident time-harmonic electromagnetic field, the electric and magnetic surface current densities induced on the
exterior side of its surface produce an electromagnetic field that cancels the incident field throughout the
interior region   of the   object. This theorem is a cornerstone of research on electromagnetic scattering,  and
particularly of the extended boundary condition
method (EBCM)
\cite{Lakh2018} since its inception in 1965 \cite{Waterman1965}.
Since a bilinear expansion of the free-space dyadic Green function is known \cite{Waterman1965,Waterman1969EM}, the EBCM
can be used to investigate   scattering by an
 object   composed of any linear homogeneous medium for which  a basis exists
 to represent the electromagnetic field therein. This requirement can be satisfied
 by  isotropic dielectric-magnetic mediums,    isotropic chiral mediums, and
 orthorhombic dielectric-magnetic mediums with gyrotropic magnetoelectric properties \cite{FarLakG4}.
 In addition,
bases have been numerically synthesized for certain gyrotropic dielectric-magnetic mediums \cite{Lin2004,Schmidt,Zouros}.
The EBCM literature continues to expand as time marches on \cite{Lakh2018,Doicu2019,Alkhoori3,Kahnert2020}

Historically, the situation has been markedly different in electrostatics.  Green's second identity
 yields an expression \cite{Jackson,Colton} that was used by Farafonov  \cite{Farafonov2014} in 2014
 to formulate the EBCM for
the perturbation (i.e., ``scattering")  of a source electric potential (the electrostatic counterpart of  an
 incident time-harmonic electromagnetic field)
by a 3D  object composed of a homogeneous isotropic medium.  In the Farafonov
formulation, Green's second identity is applied separately to the exterior and  interior regions. The potentials in the
two regions are certainly different but, as
the  Green function
\begin{equation}
\label{gf-def}
G(\#r,\#r^\prime)=\frac{1}{4\pi\vert\#r-\#r^\prime\vert}
\end{equation}
for the Poisson equation is applicable to both regions, there is no need to set up the electrostatic counterpart
of the Ewald--Oseen extinction theorem.
Eigenfunctions of the Laplace equation are used
in   infinite-series representations of the source potential everywhere, the perturbation potential in the exterior region,
as well as the  potential induced inside the isotropic dielectric object,
since
the right side of Eq.~(\ref{gf-def})
 has a bilinear expansion \cite{Morse}
in terms of those   eigenfunctions.
All three series are suitably terminated and a transition matrix is obtained to characterize the perturbation
of the source potential by the object. Although the Farafonov formulation does not require the object to be axisymmetric,
 it has been used only for such objects \cite{FU2015,FIUP2016,Majic2017}.

Our objective in this paper is to generalize the EBCM for electrostatic problems in which the
perturbing 3D object is composed of a homogeneous anisotropic dielectric medium whose  permittivity
dyadic is positive definite. We use
 Green's second identity \textit{only} in the exterior region  and deduce the electrostatic
 counterpart of the Ewald--Oseen extinction theorem therefrom. As the Laplace and   Poisson equations apply in the
 exterior region, our representations of the source and perturbation potentials are the same as in the Farafonov
 formulation. But,
the Laplace equation is inapplicable in any region
 occupied by a homogeneous anisotropic dielectric medium. Hence,  
an affine transformation of space  in the Gauss equation
for the electric field is implemented in order to generate a basis
for representing the electric potential induced inside the object \cite{LTA-1}.  This basis is premised on the
eigenfunctions of the Laplace equation in the spherical coordinate system.
There is no requirement of axisymmetry in our formulation, as is explicitly demonstrated with numerical results for perturbation
by ellipsoids.

The plan of this paper is as follows. Section~\ref{bvp} describes the boundary-value problem, Sec.~\ref{bvp-ie} contains the derivation
of integral equations underlying the EBCM for
electrostatics, Sec.~\ref{ebcm} describes the formulation of the transition matrix that relates the expansion coefficients
of the perturbation potential to those of the source potential,
and Sec.~\ref{nrd}   presents and discusses illustrative numerical results.  The paper concludes in Sec.~\ref{conc}
with remarks pertaining to future work.

\section{Boundary-Value Problem} \label{bvp}

\begin{figure}[h]
 \centering
\includegraphics[width=0.4\linewidth]{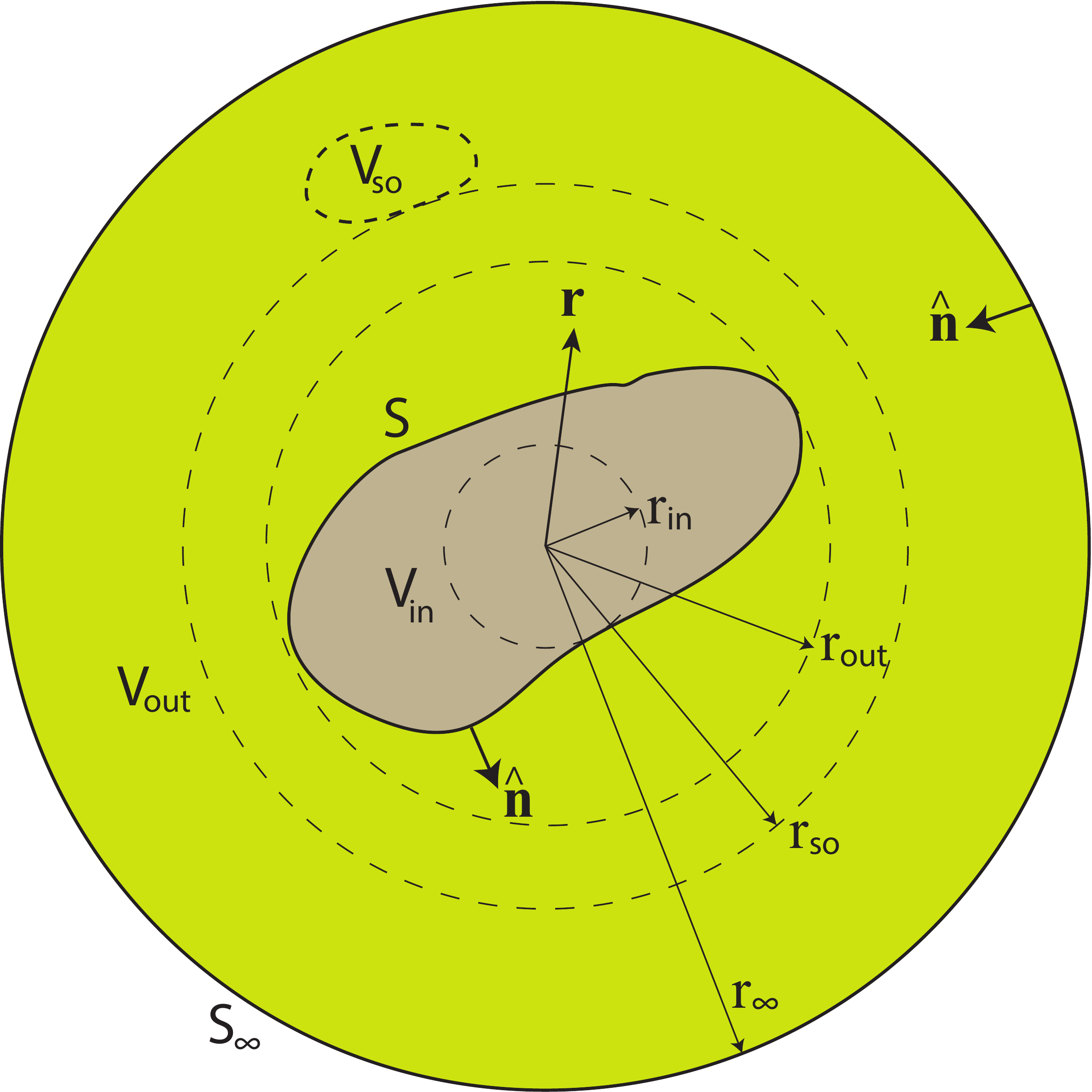}
\caption{Schematic of the boundary-value problem. The region $\Vout$ is vacuous and the region $\Vin$ is filled with an anisotropic dielectric
material of relative permittivity dyadic ${\underline{\underline{\eps}}_{\,{\rm rel}}}$, given by  Eqs.~(\ref{eps-def}) and (\ref{A-def}).}
\label{schematic}
\end{figure}

Let the    region $\Vin$ be the interior of the surface $\So$, whereas the region
$\Vout$ be bounded by  the surfaces $\So$ and $\Sinf$, as shown in Fig.~\ref{schematic}.
The coordinate
system has its origin lying inside $\Vin$, the sphere with the origin as its center and inscribed in $\Vin$
has radius $\rin$, the sphere circumscribing $\Vin$
has radius $\rout$, and $\Sinf$ is a sphere of radius $\rinf$. The surface $S$ is described by
a continuous and once-differentiable function.

The electric potential $\Phi (\#r)$ satisfies the Poisson
equation
\begin{equation}
\label{Poisson}
\nabla^2\Phi (\#r)=-\epso^{-1}\,\rho(\#r)\,,\quad \#r\in\Vout\,,
\end{equation}
in the region $\Vout$, with
  $\epso$ as the permittivity of free space (i.e., vacuum) and
 the   charge density $\rho(\#r)$ being non-zero only for $\#r\in\Vso\subset\Vout$. Every point
 in $\Vso$ is at least a distance $\rso$ away from the origin, as shown in Fig.~\ref{schematic}.

 The charge-free region
$\Vin$ is filled with an anisotropic dielectric medium  with permittivity dyadic
\begin{equation} \label{eps-def}
\=\eps
=\epso\epsrel = \epso\epsr\,\=A\.\=A\,,
\end{equation}
wherein the scalar $\epsr>0$ and
 the diagonal dyadic
\begin{equation}
\label{A-def}
\={A}= \alphax^{-1}\,\ux \ux + \alphay^{-1} \,\uy \uy + \uz \uz\,
\end{equation}
contains the anisotropy parameters $\alphax>0$ and $\alphay>0$.  
The constitutive principal axes
of this medium are thus parallel to $\ux$, $\uy$, and $\uz$.

Any real symmetric dyadic
can be written in the form of Eqs.~(\ref{eps-def}) and (\ref{A-def}) by virtue of the principal axis theorem \cite{Charnow,Strang}.
Anisotropic materials exist in nature \cite{Auld}, and homogenizable composite materials of this kind can be engineered
 by properly dispersing dielectric fibers  in some host isotropic dielectric material  \cite{MAEH}.

%%%%%%%%%%%%%%%%  Figure 1 begins %%%%%%%%%%%%%%%%%%%%
%%%%%%%%%%%%%%%%  Figure 1 ends %%%%%%%%%%%%%%%%%%%%

\section{Integral Equations}\label{bvp-ie}
Application of Green's second identity to the exterior region $\Vout$  yields \cite{Smythe,Jackson}
\begin{eqnarray}
\nonumber
&&\iiint_{\Vout}\,\Phi(\#r^\prime)\delta(\#r-\#r^\prime)\, d^3r^\prime
\\[5pt]
\nonumber
&&\quad
=\epso^{-1}\iiint_{\Vso}\rho(\#r^\prime) G(\#r,\#r^\prime)\, d^3r^\prime
\\[5pt]
\nonumber
&&\qquad
+\iint_{\So\cup\Sinf}\,\left[\Phi\plus(\#r^\prime)\nrp\.{\nabla^\prime} G(\#r,\#r^\prime)\right.
 \\[5pt]
&&\qquad\quad
\left.-G(\#r,\#r^\prime)\nrp\.{\nabla^\prime}\Phi\plus(\#r^\prime)\right] d^2r^\prime\,,
\label{FirstIE}
\end{eqnarray}
where $\Phi\plus(\#r^\prime)$ is the value of $\Phi(\#r^\prime)$ on the exterior side of $\So$,
$\delta(\#r-\#r^\prime)$ is the Dirac delta, and the unit normal vector $\nr$  at $\#r\in\So\cup\Sinf$   points into $\Vout$.

For finite $r$ and with $\Sinf$ taken to be a spherical surface of radius   $\rinf$,  the potential
$\Phi(\#r^\prime)$ at $\#r^\prime\in\Sinf$ drops off as $1/r^\prime$ as $\rinf\to\infty$. Accordingly, the integral on $\Sinf$
in Eq.~(\ref{FirstIE}) vanishes. Furthermore, let us define the source potential
\begin{equation}
\Phi\inc(\#r)=\epso^{-1}\iiint_{\Vso}\rho(\#r^\prime) G(\#r,\#r^\prime) d^3r^\prime\,,
\end{equation}
because it exists everywhere when $\Vin$ is vacuous (just like $\Vout$ actually is).
Equation~(\ref{FirstIE}) then delivers  the twin equations
\begin{subequations}
\label{SecondIE}
\begin{eqnarray}
\nonumber
&& 0
=\Phi\inc(\#r)
+\iint_{\So }\,\left[\Phi\plus(\#r^\prime)\nrp\.{\nabla^\prime} G(\#r,\#r^\prime)\right.
\\[5pt]
&&\qquad\left.-G(\#r,\#r^\prime)\nrp\.{\nabla^\prime}\Phi\plus(\#r^\prime)\right] d^2r^\prime\,,\quad
\#r\in\Vin\,,
\label{SecondIE-a}
\end{eqnarray}
and
\begin{eqnarray}
\nonumber
&& \Phi(\#r)
=\Phi\inc(\#r)
+\iint_{\So }\,\left[\Phi\plus(\#r^\prime)\nrp\.{\nabla^\prime} G(\#r,\#r^\prime)\right.
\\[5pt]
&&\qquad\left.-G(\#r,\#r^\prime)\nrp\.{\nabla^\prime}\Phi\plus(\#r^\prime)\right] d^2r^\prime\,,\quad
\#r\in\Vout\,.
\label{SecondIE-b}
\end{eqnarray}
\end{subequations}

Equation~(\ref{SecondIE-a}) is the electrostatic equivalent of the Ewald--Oseen extinction
theorem \cite{Ewald1916,Oseen1915,Lakh2018}, and is the cornerstone of our EBCM formulation.
It is convenient to rewrite it as
\begin{subequations}
\begin{eqnarray}
\nonumber
\Phi\inc(\#r)
&=&-\iint_{\So }\,\left[\Phi\plus(\#r^\prime)\nrp\.{\nabla^\prime} G(\#r,\#r^\prime)\right.
\\[5pt]
&&\quad\left.
-G(\#r,\#r^\prime)\nrp\.{\nabla^\prime}\Phi\plus(\#r^\prime)\ris d^2r^\prime\,,\quad
\#r\in\Vin\,.
\label{ThirdIE-a}
\end{eqnarray}
According to Eq.~(\ref{SecondIE-b})
the electric potential in $\Vout$ has two components,
one of which is the source potential $\Phi\inc(\#r)$ and the other is the perturbation potential $\Phi\sca(\#r)$
due to the region $\Vin$ being   non-vacuous \cite{Jackson}; indeed, $\Phi\sca(\#r)\equiv0$ if $\Vin$ is vacuous (just like
$\Vout$). Hence,
Eq.~(\ref{SecondIE-b})  yields
\begin{eqnarray}
\nonumber
&&
\Phi\sca(\#r)
= \iint_{\So }\,\left[\Phi\plus(\#r^\prime)\nrp\.{\nabla^\prime} G(\#r,\#r^\prime)\right.
\\[5pt]
&&\qquad\left.
-G(\#r,\#r^\prime)\nrp\.{\nabla^\prime}\Phi\plus(\#r^\prime)\ris d^2r^\prime\,,\quad
\#r\in\Vout\,.
\label{ThirdIE-b}
\end{eqnarray}
\end{subequations}

With the usual requirement
of the electric field having a finite magnitude everywhere on $\So$, the
potential must be continuous across that interface; hence,
\begin{subequations}
\begin{equation}
\label{bc1}
\Phi\plus(\#r) =\Phi\minus(\#r)\,,
\quad \#r\in\So\,,
\end{equation}
where $\Phi\minus(\#r^\prime)$ is the value of $\Phi(\#r^\prime)$ on the interior side of $\So$.
Likewise, with the assumption of $\So$ being charge-free, the normal component of the
electric displacement must be continuous across that surface; hence,
\begin{equation}
\label{bc2}
\nr\.\nabla\Phi\plus(\#r)= \nr\.\epsrel\.\nabla\Phi\minus(\#r)\,,
\quad \#r\in\So\,.
\end{equation}
\end{subequations}

Substitution of Eqs.~(\ref{bc1}) and (\ref{bc2}) in Eqs.~(\ref{ThirdIE-a}) and (\ref{ThirdIE-b}) delivers
\begin{eqnarray}
\nonumber
&&
\left.\begin{array}{c}
\Phi\inc(\#r)\\
\Phi\sca(\#r)
\end{array}\right\}
=\mp\iint_{\So }\,\left[\Phi\minus(\#r^\prime)\,\nrp\.{\nabla^\prime} G(\#r,\#r^\prime)\right.
\\[5pt]
&&\qquad
\left.
-G(\#r,\#r^\prime)\,\nrp\.\epsrel\.{\nabla^\prime}\Phi\minus(\#r^\prime)\right] d^2r^\prime\,,\quad
\left\{\begin{array}{l}
\#r\in\Vin\,,
\\
\#r\in\Vout\,.
\end{array}
\right.
\label{FourthIE}
\end{eqnarray}
Since $\Phi\minus(\#r)$ at $\#r\in\So$ is nothing but the internal potential $\Phi\ind(\#r)$ evaluated
on the interior side of $\So$, we finally get
\begin{eqnarray}
\nonumber
&&
\left.\begin{array}{c}
\Phi\inc(\#r)\\
\Phi\sca(\#r)
\end{array}\right\}
=\mp\iint_{\So }\,\left[\Phi\ind(\#r^\prime)\,\nrp\.{\nabla^\prime} G(\#r,\#r^\prime)\right.
\\[5pt]
&&\qquad
\left.
-G(\#r,\#r^\prime)\,\nrp\.\epsrel\.{\nabla^\prime}\Phi\ind(\#r^\prime)\right] d^2r^\prime\,,\quad
\left\{\begin{array}{l}
\#r\in\Vin\,,
\\
\#r\in\Vout\,.
\end{array}
\right.
\label{FifthIE}
\end{eqnarray}
Knowing $\Phi\inc$,
we first have to find $\Phi\ind$ everywhere in $\Vin$ after choosing $\#r\in\Vin$ in Eqs.~(\ref{FifthIE}). Thereafter,
knowing $\Phi\ind$,
we can  find $\Phi\sca$ everywhere in $\Vout$
after choosing  $\#r\in\Vout$ in Eqs.~(\ref{FifthIE}). Thus, the electrostatic counterpart
of the Ewald--Oseen formulation is a crucial ingredient in our EBCM formulation, which follows the style of the Waterman formulation
for electromagnetics \cite{Waterman1965,Waterman1969EM}.

\section{Extended Boundary Condition Method}\label{ebcm}
\subsection{Source and perturbation potentials}
The solutions of the Laplace equation in the spherical coordinate system being known \cite{Morse},
the source potential can be represented as
\begin{equation}
\label{Phi-inc}
\Phi\inc(\#r)= \sum_{s\in\left\{e,o\right\}}^{}\sum_{n=0}^{\infty}\sum_{m=0}^{n}\les \Emn
\calA\smn \,r^n  \, Y\smn(\theta,\phi)\ris\,,\quad r<\rso\,.
\end{equation}
Here, the expansion coefficients $\calA\smn$ are supposed to be known,
\begin{equation}
\Emn =  \left(2-\delta_{m0}\right)\frac{2n+1}{4\pi}\,\frac{(n-m)!}{(n+m)!}\,
\end{equation}
is a normalization factor with $\delta_{mm^\prime}$ as the Kronecker delta,
and
the  spherical harmonics
\begin{equation}
\left.\begin{array}{l}
Y\emn(\theta,\phi) = P_n^m(\cos\theta) \cos(m\phi)
\\[5pt]
Y\omn(\theta,\phi) = P_n^m(\cos\theta) \sin(m\phi)
\end{array}\right\}\,
\end{equation}
involve the associated Legendre function $ P_n^m(\cos\theta)$  \cite{Morse}.
The definitions of the spherical harmonics mandate that $\calA_{\rm o0n}= 0\,\forall{n\in[0,\infty)}$.
Likewise, the perturbation potential can be represented as
\begin{equation}
\label{Phi-sca}
\Phi\sca(\#r)= \sum_{s\in\left\{e,o\right\}}^{}\sum_{n=0}^{\infty}\sum_{m=0}^{n}\les \Emn
\calB\smn\, r^{-(n+1)} \,  Y\smn(\theta,\phi)\ris \,,\quad r \geq\rout\,,
\end{equation}
 with unknown expansion coefficients $\calB\smn$, with $\calB_{\rm o0n}= 0\,\forall{n\in[0,\infty)}$.

 \subsection{Algebraic equations}
The Green function $G(\#r,\#r^\prime)$
 has the bilinear expansion \cite{Morse}
 \begin{eqnarray}
 \nonumber
 G(\#r,\#r^\prime)&=&  \displaystyle{
 \sum_{s\in\left\{e,o\right\}}^{}\sum_{n=0}^{\infty}\sum_{m=0}^{n}\left[ \frac{\Emn}{2n+1}
Y\smn(\theta,\phi)Y\smn(\theta^\prime,\phi^\prime)\right.}
\\[5pt]
&&
\left.
\lec
\begin{array} {l}
r^n (r^\prime)^{-(n+1)}
\\[6pt]
r^{-(n+1)} (r^\prime)^{n}
\end{array}\ric
\right]
\,,
\quad
\left\{
\begin{array}{l}
r < r^\prime\,,
\\[6pt]
r> r^\prime\,.
\end{array}
\right.
\end{eqnarray}
Now, we     use   the orthogonality relationships of the spherical harmonics on spherical surfaces.
First, let   the surface $r= \ain <\rin$ be chosen for applying Eq.~(\ref{FifthIE})$_{\rm top}$. The source potential on the left side
of this equation can be replaced
by the series on the right side of Eq.~(\ref{Phi-inc}). The bilinear
expansion of  $G(\#r,\#r^\prime)$ for $r < r^\prime$ has to be used on the right side of  Eq.~(\ref{FifthIE})$_{\rm top}$, since $r=\ain$ and $\#r^\prime\in\So$. After multiplying both sides
of the resulting equation by $Y\smnp(\theta,\phi)$ and integrating over the surface $r= \ain$, we obtain
\begin{subequations}
\begin{eqnarray}
\nonumber
&&\calA\smn=-\,\frac{1}{2n+1}
\iint_{\So }\,\lec\Phi\ind(\#r^\prime)\,\nrp\.{\nabla^\prime} \les(r^\prime)^{-(n+1)}\,Y\smn(\theta^\prime,\phi^\prime)\ris\right.
\\[7pt]
&&\left.\qquad
-(r^\prime)^{-(n+1)}\,Y\smn(\theta^\prime,\phi^\prime)\,\nrp\.\epsrel\.{\nabla^\prime}\Phi\ind(\#r^\prime)\ric d^2r^\prime\,.
\label{SixthIE-top}
\end{eqnarray}

Second,  we choose   the surface $r= \aout >\rout$   for applying Eq.~(\ref{FifthIE})$_{\rm bot}$. The perturbation potential on the left side
of this equation can be replaced
by the series on the right side of Eq.~(\ref{Phi-sca}). On the right side of  Eq.~(\ref{FifthIE})$_{\rm bot}$  the bilinear
expansion of  $G(\#r,\#r^\prime)$ for $r > r^\prime$ must be used, since $r=\aout$ and $\#r^\prime\in\So$. Then, after multiplying both sides
of the resulting equation by $Y\smnp(\theta,\phi)$ and integrating over the surface $r= \aout$, we get
\begin{eqnarray}
\nonumber
\calB\smn&=& \frac{1}{2n+1}
\iint_{\So }\,\lec
\Phi\ind(\#r^\prime)\,\nrp\.{\nabla^\prime}
\les
(r^\prime)^{n}\,Y\smn(\theta^\prime,\phi^\prime)
\ris
\right.
\\[7pt]
&&\left.\qquad
-(r^\prime)^{n}\,Y\smn(\theta^\prime,\phi^\prime)\,\nrp\.\epsrel\.{\nabla^\prime}\Phi\ind(\#r^\prime)\ric d^2r^\prime\,.
\label{SixthIE-bot}
\end{eqnarray}
\end{subequations}

The integral equations (\ref{SixthIE-top}) and (\ref{SixthIE-bot}) incorporate the boundary conditions (\ref{bc1})
and (\ref{bc2}) prevailing on the surface $\So$ of the perturbing object $\Vin$, but the orthogonalities of the
spherical harmonics were not applied on that surface. This is the reason for ``extended boundary condition" in the
name of EBCM \cite{Waterman1965,Lakh2018}.

\subsection{Internal potential}

The   internal  potential $\Phi\ind(\#r)$ satisfies the  equation
\begin{equation}
\label{eq18}
\nabla\.\lec\=\eps\.\les\nabla \Phi\ind(\#r)\ris\ric = 0\,,\quad \#r\in  \Vin\,,
\end{equation}
which emerges from the Gauss equation for the electric field. Equation~(\ref{eq18}) is not the Laplace equation,
although it does
simplify to the Laplace equation   when $\alphax=\alphay=1$ (i.e., for an isotropic medium
\cite{Farafonov2014,FU2015,FIUP2016,Majic2017}).

After applying a bijective affine transformation of space that maps a sphere into an ellipsoid,
the solution of Eq.~(\ref{eq18})
can be written as the series \cite{LTA-1}
\begin{equation}
\label{Phi-int}
\Phi\ind(\#r)= \sum_{s\in\left\{e,o\right\}}^{}\sum_{n=0}^{\infty}\sum_{m=0}^{n}\les
\calC\smn\,  Z\smn(\#r)\ris\,,\quad \#r\in\Vin\,,
\end{equation}
where
\begin{equation}
\left.\begin{array}{l}
Z\emn(\#r) = \vert\=A^{-1}\.\#r\vert^n\,P_n^m \les\frac{r}{\vert\=A^{-1}\.\#r\vert} \cos\theta\ris
\cos\les m \tan^{-1}\left(\frac{\alphay}{\alphax}\,\tan\phi\right)\ris
\\[8pt]
Z\omn(\#r) = \vert\=A^{-1}\.\#r\vert^n\,P_n^m \les\frac{r}{\vert\=A^{-1}\.\#r\vert} \cos\theta\ris
\sin\les m \tan^{-1}\left(\frac{\alphay}{\alphax}\,\tan\phi\right)\ris
\end{array}\right\}\,.
\end{equation}
Since the basis formed by the functions
$Z\smn(\#r)$
 is constructed by a spatial transformation of the
eigenfunctions of the Laplace equation in the spherical coordinate system,
care must be taken to ensure that the angle
\begin{subequations}
\begin{equation}
\theta_q=\cos^{-1}\left(\frac{r}{\vert\=A^{-1}\.\#r\vert} \cos\theta\right)\,
\end{equation}
 lies in the same quadrant as $\theta$, and the angle
\begin{equation}
\phi_q=\tan^{-1}\left(\frac{\alphay}{\alphax}\tan\phi\right)\,
\end{equation}
\end{subequations}
lies in the same
quadrant as $\phi$.

\subsection{Transition matrix}
Substitution of Eq.~(\ref{Phi-int}) in Eq.~(\ref{SixthIE-top}) delivers the algebraic equation
\begin{eqnarray}
\calA\smn&=&-\,  \sum_{s^\prime\in\left\{e,o\right\}}^{}\sum_{n^\prime=0}^{\infty}\sum_{m^\prime=0}^{n^\prime}\left(
\calQ^{(1)}
\double\,\calC\smnp \right)\,,
\label{SeventhIE-top}
\end{eqnarray}
where the surface integral
\begin{eqnarray}
\nonumber
&&\calQ^{(1)}
\double= \frac{1}{2n+1} \iint_{\So }\,\lec \nrp\.{\nabla^\prime} \les(r^\prime)^{-(n+1)}\,Y\smn(\theta^\prime,\phi^\prime)\ris Z\smnp(\#r^\prime) \right.
\\[7pt]
&&
\left.
-(r^\prime)^{-(n+1)}\,Y\smn(\theta^\prime,\phi^\prime)\,\nrp\.\epsrel\.{\nabla^\prime}Z\smnp(\#r^\prime)\,\ric d^2r^\prime \,.
\label{Q1-def}
\end{eqnarray}
Likewise, substitution of Eq.~(\ref{Phi-int}) in Eq.~(\ref{SixthIE-bot}) leads to
\begin{eqnarray}
\calB\smn&=&   \sum_{s^\prime\in\left\{e,o\right\}}^{}\sum_{n^\prime=0}^{\infty}\sum_{m^\prime=0}^{n^\prime}\left(
\calQ^{(3)}
\double\,
\calC\smnp \right)\,,
\label{SeventhIE-bot}
\end{eqnarray}
where the surface integral
\begin{eqnarray}
\nonumber
\calQ^{(3)}
&&\double= \frac{1}{2n+1}
\iint_{\So }\,\lec \nrp\.{\nabla^\prime} \les(r^\prime)^{n}\,Y\smn(\theta^\prime,\phi^\prime)\ris\,Z\smnp(\#r^\prime)\right.
\\[7pt]
&&\left.
-(r^\prime)^{n}\,Y\smn(\theta^\prime,\phi^\prime)\,\nrp\.\epsrel\.{\nabla^\prime}Z\smnp(\#r^\prime)\ric d^2r^\prime \,.
\label{Q3-def}
\end{eqnarray}

After   the indexes $n$ and $n^\prime$ in the foregoing equations are restricted to the range $[0,N]$, Eq.~(\ref{SeventhIE-top})
leads to the matrix equation (in abbreviated notation)
\begin{equation}
\label{Q1-eq}
\calA =-\calQ^{(1)}\.\calC\,,
\end{equation}
whereas Eq.~(\ref{SeventhIE-bot})
yields   the matrix equation
\begin{equation}
\label{Q3-eq}
\calB=\calQ^{(3)}\.\calC\,.
\end{equation}
Then,
\begin{equation} \label{B-eq}
\calB=\calT\.\calA\,,
\end{equation}
where the transition matrix
\begin{equation} \label{T-eq}
\calT=-\calQ^{(3)}\.\left(\calQ^{(1)}\right)^{-1}
\end{equation}
relates the expansion coefficients of the perturbation potential to those of the source potential.
Very importantly, this matrix does not depend on the source potential.

\subsection{Asymptotic expression for perturbation potential}

The perturbation potential defined in Eq.~(\ref{Phi-sca}) can be written as
\begin{subequations}
\begin{eqnarray}
\nonumber
&&\Phi\sca(\#r)
=   \frac{1}{r} \,\fscaone +
\frac{1}{{r^2}}\,\fscatwo(\theta,\phi)
\\[5pt]
\label{Phi-sca-1}
&& +\lim_{N\to\infty}
\sum_{s\in\left\{e,o\right\}}^{}\sum_{n=2}^{N}\frac{1}{r^{n+1}}
\sum_{m=0}^{n}\les  \Emn
\calB\smn  \,  Y\smn(\theta,\phi)\ris \,,
\quad r \geq \rout\,,
\end{eqnarray}
where
\begin{equation}
\label{fpert1-def}
\fscaone=\frac{1}{4 \pi  } \calB_{\rm e00}
\end{equation}
and
\begin{equation}
\label{fpert2-def}
\fscatwo(\theta,\phi)=\frac{3}{4 \pi  } \les\calB_{\rm e01} \cos\theta +2 \left(\calB_{\rm e11}\cos\phi + \calB_{\rm o11}\sin\phi\right)\sin\theta\ris\,.
\end{equation}
\end{subequations}
Thus, far away
from the object,   the
perturbation potential can be asymptotically stated as
\begin{equation}
\label{Phisca-far}
\Phi\sca(\#r)=  \frac{1}{r} \,\fscaone +  \frac{1}{{r^2}}\,\fscatwo(\theta,\phi) +\mathcal{O}\left(\frac{1}{r^3}\right),\,\,\,r\rightarrow \infty\,,
\end{equation}
and it is determined only by the four
 perturbational-potential coefficients  $\calB_{\rm e00}$,  $\calB_{\rm e01}$, $\calB_{\rm e11}$, and $\calB_{\rm o11}$.

We have explicitly retained the two lowest-order terms in  Eq.~(\ref{Phisca-far}), because $\fscaone$ is not always nonzero. For instance,
$\calB_{\rm e00}=0$ when the source potential varies linearly in a fixed direction and
object is axisymmetric as well as composed of an isotropic medium \cite{Farafonov2014,FU2015,Majic2017}.
Likewise, $\calB_{\rm e00}=0$
when   the object is a sphere and the source is either a point charge or a point dipole, so that
$\fscaone=0$ and $\Phi\sca(\#r)\propto r^{-2}$ as $r\rightarrow\infty$ \cite{LTA-1}. 
More generally, the distance after
which $\fscaone$ dominates $\fscatwo$
can depend on the direction, because $\fscaone$ is independent of $\theta$ and $\phi$
whereas $\fscatwo$ is not. Indeed,
 the first term on the right side of Eq.~(\ref{fpert2-def}) does not exist for $\theta=\pi/2$,
whereas the second and third terms on the right side of the same equation are absent for $\theta\in\lec0,\pi\ric$.

\section{Numerical Results and Discussion}\label{nrd}

By virtue of Eqs. (\ref{Q1-def}), (\ref{Q3-def}), (\ref{B-eq}), and (\ref{T-eq}), the perturbation potential must depend on (i) the geometry of the perturbing object  described via $\So$; (ii) the composition of the object, quantitated by $\epsrel$; and (iii) the spatial
profile of the source potential $\Phi\inc$. We present numerical results in this section to illustrate the effects
of  $\So$, $\epsrel$, and $\Phi\inc$ on the perturbation potential $\Phi\sca$.

\subsection{Preliminaries}

\subsubsection{Geometry}
We chose  $\So$ to be an ellipsoidal surface defined by the position vector
\begin{eqnarray}
&&
\nonumber
\#r(\theta,\phi)= a \, \=S \. \=U\. \=S^{-1} \.\left[\left(\ux\cos\phi+\uy\sin\phi\right)\sin\theta +\uz\cos\theta\right]\,,
\\[5pt]
&&\quad
\quad \theta\in[0,\pi]\,,\quad\phi\in[0,2\pi)\,.
\label{rs-def}
\end{eqnarray}
The rotation dyadic
\begin{subequations}
\begin{equation}
\={S}= \={R}_z(\gammas) \. \={R}_y(\betas) \. \={R}_z(\alphas)
\label{S-def}
\end{equation}
is a product of three rotation dyadics, with
\begin{eqnarray}
\nonumber
&&\={R}_z(\zeta)= (\ux \ux+ \uy \uy) \cos \zeta - (\ux \uy-\uy \ux) \sin \zeta + \uz \uz\,,
\\[5pt]
&&\qquad\quad
\zeta\in\left\{\alphas,\gammas\right\}\,,
\end{eqnarray}
and
\begin{eqnarray}
&&\={R}_y(\betas)= (\ux \ux+ \uz \uz) \cos \betas - (\uz \ux-\ux \uz) \sin \betas + \uy \uy\,.
\end{eqnarray}
\end{subequations}
Sequentially,
$\={R}_z(\alphas)$ represents a rotation by $\alphas\in [0,\pi]$ about the $z$ axis,  $\={R}_y(\betas)$ represents  a rotation by $\betas \in [0,\pi]$ about the new $y$ axis, and  $\={R}_z(\gammas)$ represents a rotation by $\gammas \in [0,\pi]$ around the new $z$ axis.
The shape dyadic is given by
\begin{equation}
\label{U-def}
\=U= \mu \, \ux\ux+ \nu \, \uy\uy + \uz\uz\,.
\end{equation}

Thus, the shape principal axes of the ellipsoidal object   are parallel to the unit vectors $\=S\.\ux$, $\=S\.\uy$, and $\=S\.\uz$.
The semi-axes of this object are given by $a\mu$ parallel to  $\=S\.\ux$,
$a\nu$ parallel to  $\=S\.\uy$, and $a$ parallel to  $\=S\.\uz$, where $\mu>0$ and $\nu>0$ are the two aspect ratios of the
ellipsoid; thus, $\rout =\max\lec{a,a\mu,a\nu}\ric$. Finally, with the
 geometric mean $\aave$ of the three semi-major axes fixed, the length
\begin{equation}
a =   {\aave}\left({\mu \nu} \right)^{-1/3}
\end{equation}
becomes a function of the aspect ratios $\mu$ and $\nu$.
Note that $a=\aave$ when $\mu = \nu = 1$ (i.e., the ellipsoid reduces to a sphere).

\subsubsection{Composition}
With  the arithmetic mean $\epsave$ of the three eigenvalues of $\epsrel$ fixed,
the relative permittivity scalar
\begin{equation}
\epsr = \frac{3 \, \epsave}{\alphax^{-2}+ \alphay^{-2} + 1}
\end{equation}
is a function of the anisotropy parameters $\alphax$ and $\alphay$.
Note that $\epsr=\epsave$ when $\alpha_x = \alpha_y = 1$ (i.e., the material becomes isotropic).

\subsubsection{Source}
Sources of two types were considered for illustrative numerical results: (i) a point-charge source and (ii) a point-dipole source.
The coefficient \cite{Jackson}
\begin{equation}
\label{Atilde-ps-p}
\calA\smn = \frac{Q}{\epso} \,\frac{1}{2n+1} \,\ro^{-(n+1)}\,Y\smn(\thetao,\phio)\,
\end{equation}
for a point charge $Q$ located at  $\bro\equiv(\ro,\thetao,\phio)$ with $\ro\ge \rout$, $\thetao \in[0,\pi]$, and $\phio \in[0,2\pi)$.
For a point dipole $\#p = p\, \hat{\#p}$ located at the same point, the coefficient \cite{Tsitsas}
\begin{equation}
\label{Atilde-ps-d}
\calA\smn =\frac{p}{\epso} \,\frac{1}{2n+1}\, \hat{\#p}  \.\nabla_{\rm o}
\les\ro^{-(n+1)}\,Y\smn(\thetao,\phio)\ris\,,
\end{equation}
where $p>0$; the unit vector
\begin{equation}
\up=  \left( \ux\cos\phip+\uy\sin\phip\right)\sin\thetap +\uz\cos\thetap
\end{equation}
contains the angles  $\thetap\in[0,\pi]$ and $\phip\in[0,2\pi)$ that define the orientation of the dipole, and $\nabla_{\rm o}(...)$ denotes the gradient with respect to $\bro$.

\subsection{Convergence}
A Mathematica\texttrademark~program was written to calculate the transition matrix $\calT$ of an anisotropic ellipsoid
for a chosen $N$.
The column vector $\calB$ containing
the
perturbation-potential coefficients was then calculated using Eq.~(\ref{B-eq}).
Also determined was the spatial profile of
  $\Phi\sca(r,\theta,\phi)$.

A convergence test was carried out with respect to $N$, by calculating the integral
\begin{equation}
I(r)=
\int_{\phi=0}^{2 \pi}
\int_{\theta=0}^{\pi}  \Phi_\text{pert}^2(r,\theta,\phi) \sin \theta \, d \theta \, d \phi\,,
\end{equation}
at diverse values of  $r \in[1,10]\,\rout$ as $N$ was incremented by unity. The iterative process of increasing $N$ was terminated when
$I(r)$ converged within a preset tolerance of \red{$1\%$} for
a specific value of $r$.  The adequate value of $N$ was higher for lower $r$, with $N= 7$ sufficient for  $r\geq 1.1\rout$.

\subsection{Code validation}
Validation of the EBCM code was done in two steps.  First, the results for the perturbation of the potential of a
point charge by a prolate spheroid (i.e., $\mu=\nu<1$) composed of an isotropic dielectric medium were compared
 with the corresponding series solution obtained using the eigenfunctions of the Laplace
 equation in the prolate-spheroidal coordinate system \cite{Redzic}.

\begin{figure}[H]
 \centering
\includegraphics[width=0.4\linewidth]{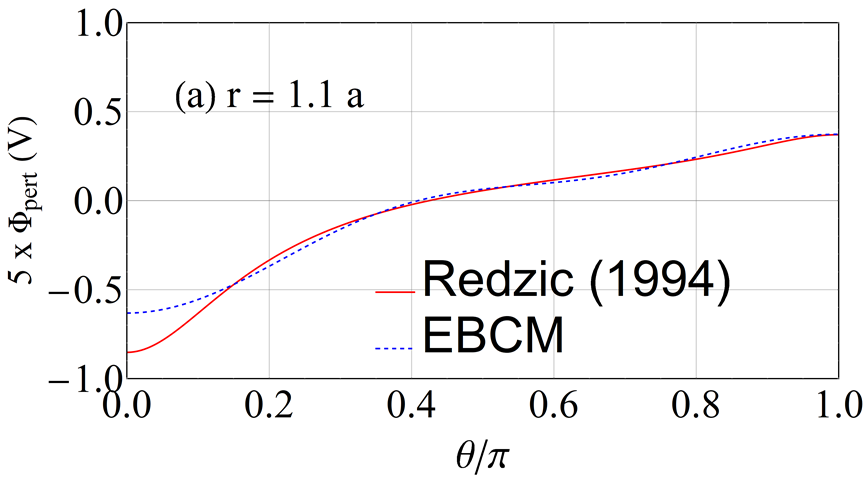}
\includegraphics[width=0.4\linewidth]{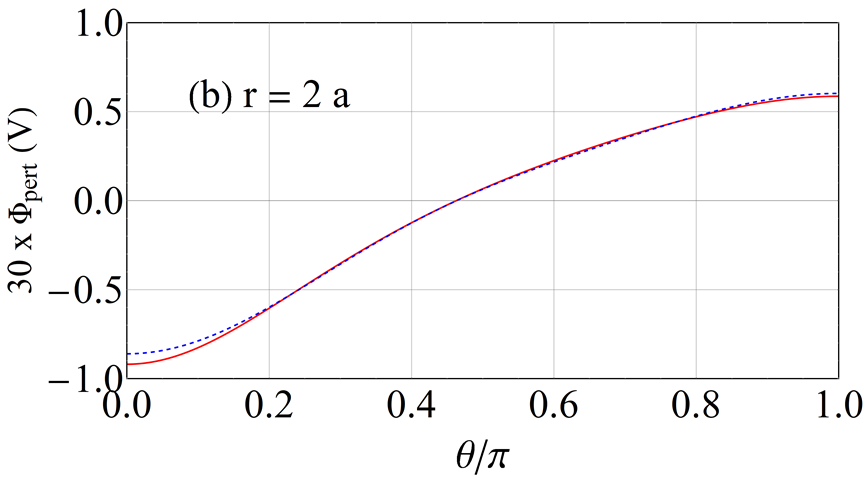} \\
\includegraphics[width=0.4\linewidth]{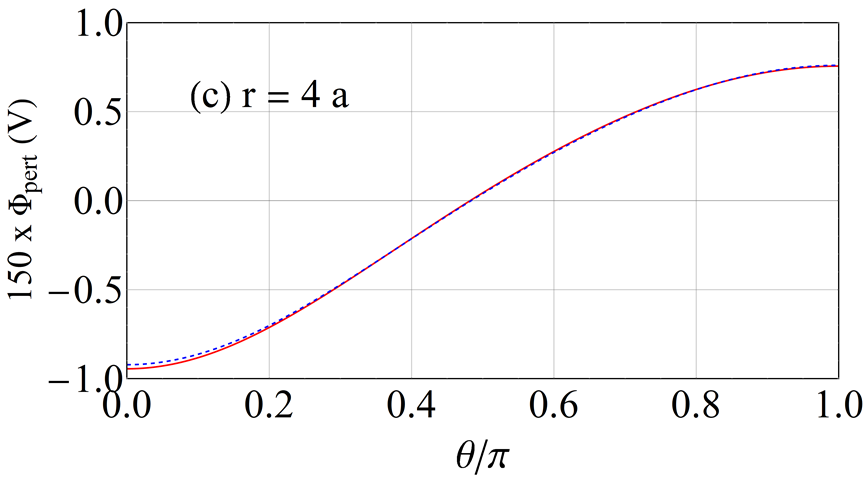}
\includegraphics[width=0.4\linewidth]{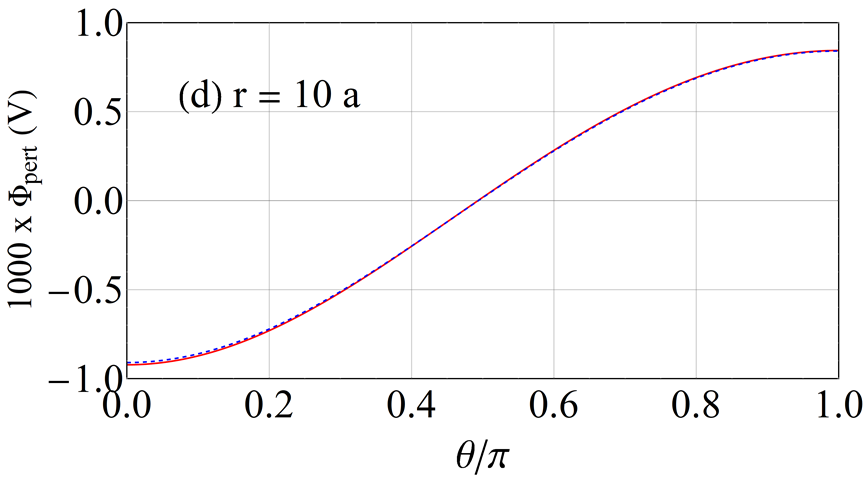}
\caption{$\Phi\sca(r,\theta,0)$ vs. $\theta$ for a prolate spheroid when $\aave = 3.82$~cm,
$\mu = \nu = 2/3$,    {$\epsave = 3$}, and $\alphax=\alphay=1$. The source is a point charge with
$Q =  \left({10^{-9}}/{36 \pi}\right)$~C located at $\ro=2a$, $\thetao=0$.
%, and $\phio=0$. 
(a) $r = 1.1a$, (b)  $r = 2a$, (c)  $r = 4a$, and (d) $r = 10a$.
}
\label{Fig2}
\end{figure}

Figure \ref{Fig2} shows plots of the   perturbation potential $\Phi\sca(r,\theta,0)$ evaluated at
$\rtilde=r/\rout \in\lec1.1,2,4,10\ric$. These plots were obtained using the series solution \cite{Redzic} and our EBCM when the source is a point charge
$Q =    \left({10^{-9}}/{36 \pi}\right)$~C located at $\ro =2\rout$ and $\thetao=0$ 
(with $\phio$ being irrelevant as $\sin\thetao=0$),
whereas the perturbing object is characterized by  $\aave = 3.82$~cm,  $\mu = \nu = 2/3$,  $\epsave = 3$, and $\alphax=\alphay=1$. Good agreement  can be seen in Fig.~\ref{Fig2}(a) between the two solutions when  $\rtilde = 1.1$, except in the neighborhood of $\theta=0$. The difference arises because the spherical harmonics used for representing $\Phi\sca$ and $\Phi\ind$ do not follow the shape of the prolate spheroid well away from the equator, an issue associated with EBCM even for time-harmonic  problems \cite{Isk1,Isk2}. However, the difference diminishes as $\rtilde$ increases. Indeed, the difference is vanishingly small for  $\rtilde = 10$ in Fig.~\ref{Fig2}(d).

%%%%%%%%%%%%%%%%  Figure 2 begins %%%%%%%%%%%%%%%%%%%%
%%%%%%%%%%%%%%%%  Figure 2 ends %%%%%%%%%%%%%%%%%%%%

Second, the same comparison was done for a sphere (i.e., $\mu=\nu=1$)
 made of an anisotropic dielectric medium, the series solution for a sphere being available using the eigenfunctions of the Laplace
 equation in the spherical coordinate system \cite{LTA-1}. The difference in values of $\Phi\sca(\#r)$ calculated using the
 two methods
 for any $r\geq{a}$ was always less than $0.001\%$.

\subsection{Numerical results for anisotropic dielectric ellipsoids}

Now we present the perturbation potential's variations with respect to $\rtilde$ and $\=S$ when the source is either a
point charge or a point dipole. For definiteness, we fixed $\aave=5$~cm, $\mu = 0.8$, $\nu = 1.2$, $\epsave=3$, $\alphax = 0.5$, and $\alphay = 1.5$. Also, the point source was located at $\ro = 2 \rout$, $\thetao = \pi/4$, and $\phio = \pi/6$.

\subsubsection{Point-charge source}\label{pcs}

\begin{figure}[H]
 \centering
\includegraphics[width=0.3\linewidth]{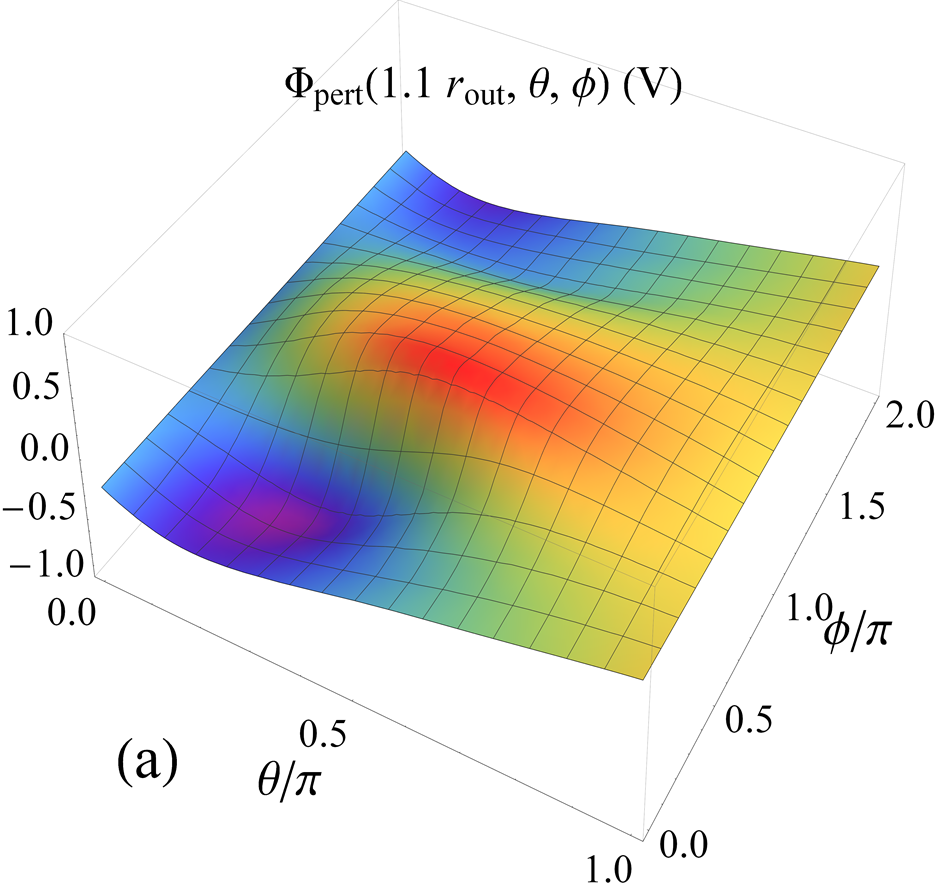}
\includegraphics[width=0.3\linewidth]{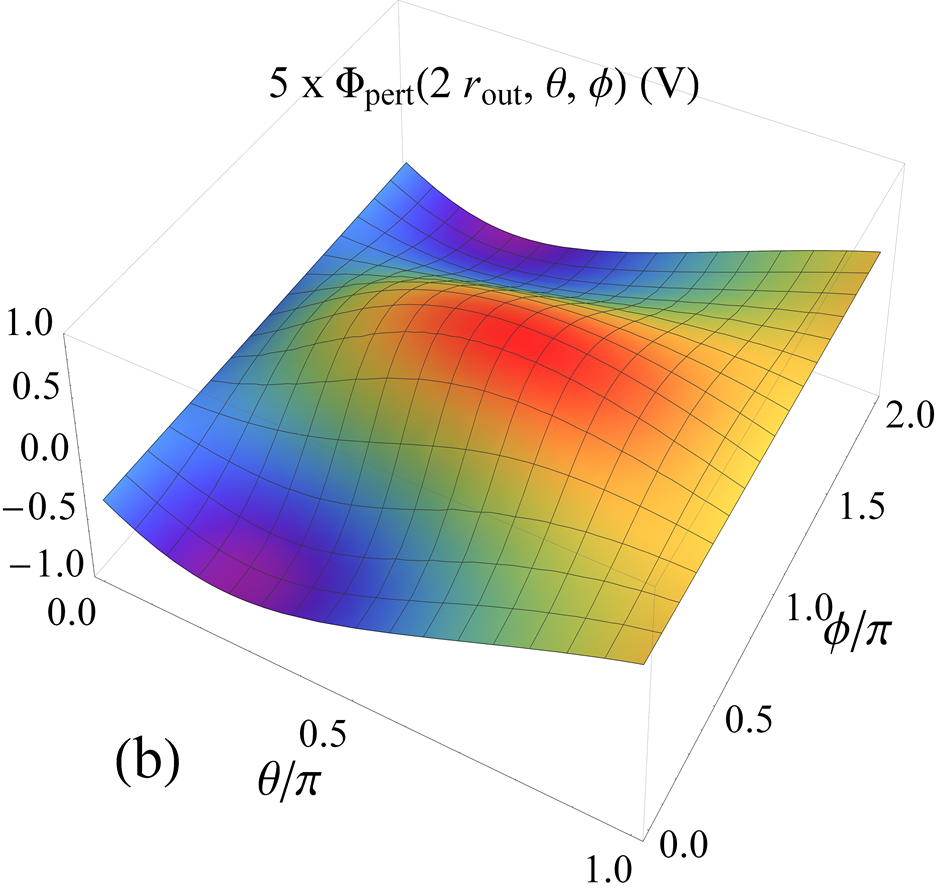} \\
\includegraphics[width=0.3\linewidth]{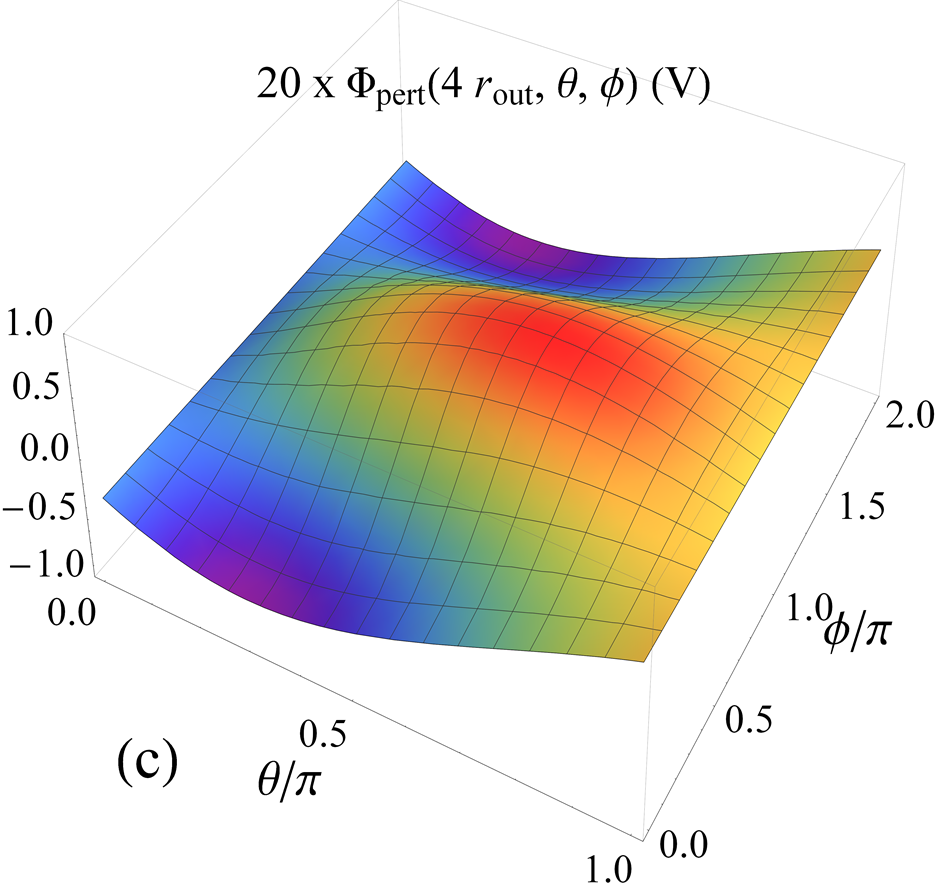}
\includegraphics[width=0.3\linewidth]{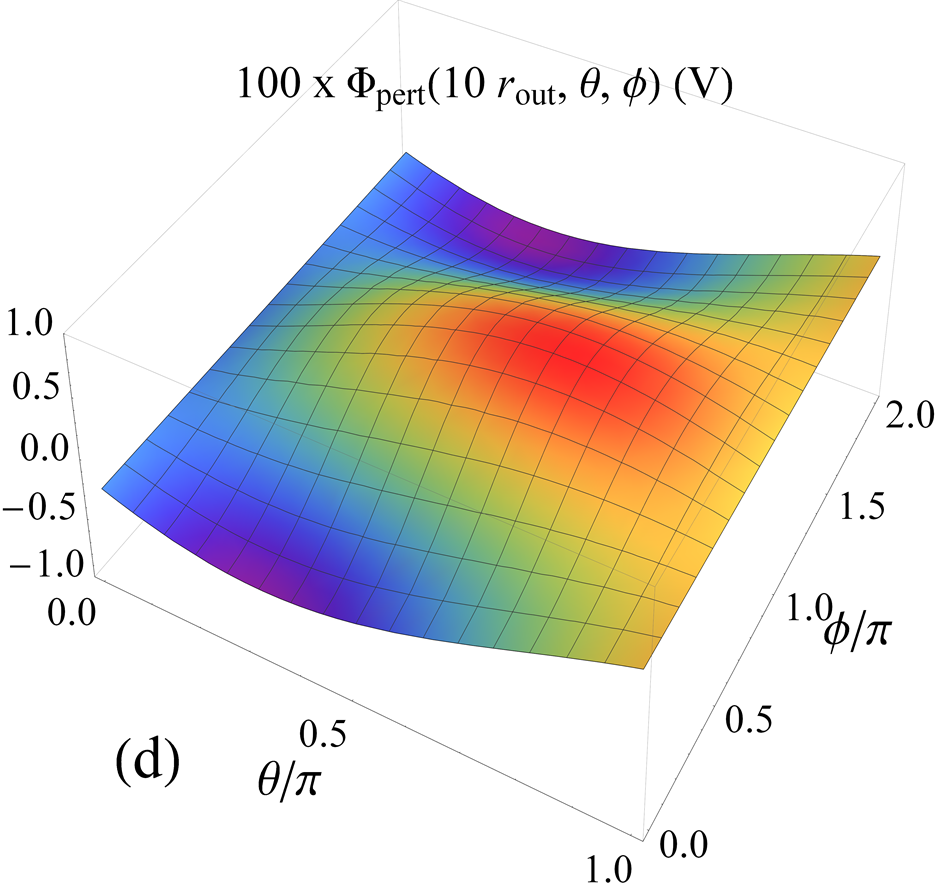}
\caption{$\Phi_\text{pert}(\rtilde\,\rout, \theta,\phi)$ versus $\theta$ and $\phi$ when $\aave=5$~cm,
$\mu = 0.8$, $\nu = 1.2$,  $\epsave=3$, $\alphax = 0.5$,   $\alphay = 1.5$, and $\alphas = \betas = \gammas = 0$. The source
is a point charge    $Q = 10^{-10}$~C  located at $\ro = 2 \rout$, $\thetao = \pi/4$, and $\phio = \pi/6$. (a) $\rtilde = 1.1$, (b) $\rtilde = 2$, (c) $\rtilde = 4$, and (d) $\rtilde = 10$. }
\label{Fig3}
\end{figure}

In order to focus on the effect of $\rtilde$ exclusively, we set $\=S = \=I$ (i.e., the shape principal axes coincide with the constitutive principal axes). Figures \ref{Fig3}(a--d) present the angular profiles of  $\Phi_\text{pert}(\rtilde\,\rout, \theta,\phi)$ for $\rtilde \in \{ 1.1, 2, 4, 10\}$ for a point-charge source with $Q = 10^{-10}$~C, when $\alphas = \betas = \gammas = 0$. The perturbation potential  $\Phi_\text{pert}(\rtilde\,\rout, \theta,\phi)$ decreases with increase of $\rtilde$, as becomes evident by comparing Figs.~\ref{Fig3}(a), \ref{Fig3}(b), \ref{Fig3}(c), and \ref{Fig3}(d). The angular profiles vary significantly for $\rtilde\leq 2$, but they
change very little as $\rtilde-2$ increases. We have observed through diverse computations
(results not shown) that  the foregoing observation is valid even when $\ro$ exceeds $2 \rout$.

%%%%%%%%%%%%%%%%  Figure 3 begins %%%%%%%%%%%%%%%%%%%%
%%%%%%%%%%%%%%%%  Figure 3 ends %%%%%%%%%%%%%%%%%%%%

\begin{figure}[H]
 \centering
\includegraphics[width=0.3\linewidth]{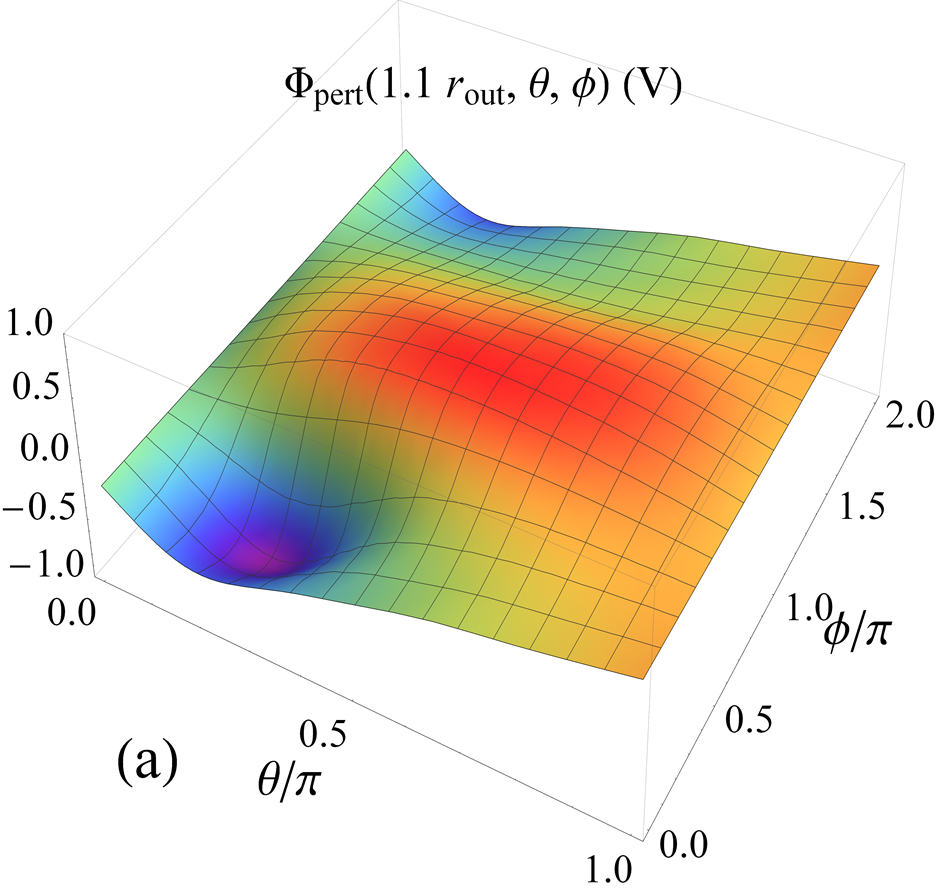}
\includegraphics[width=0.3\linewidth]{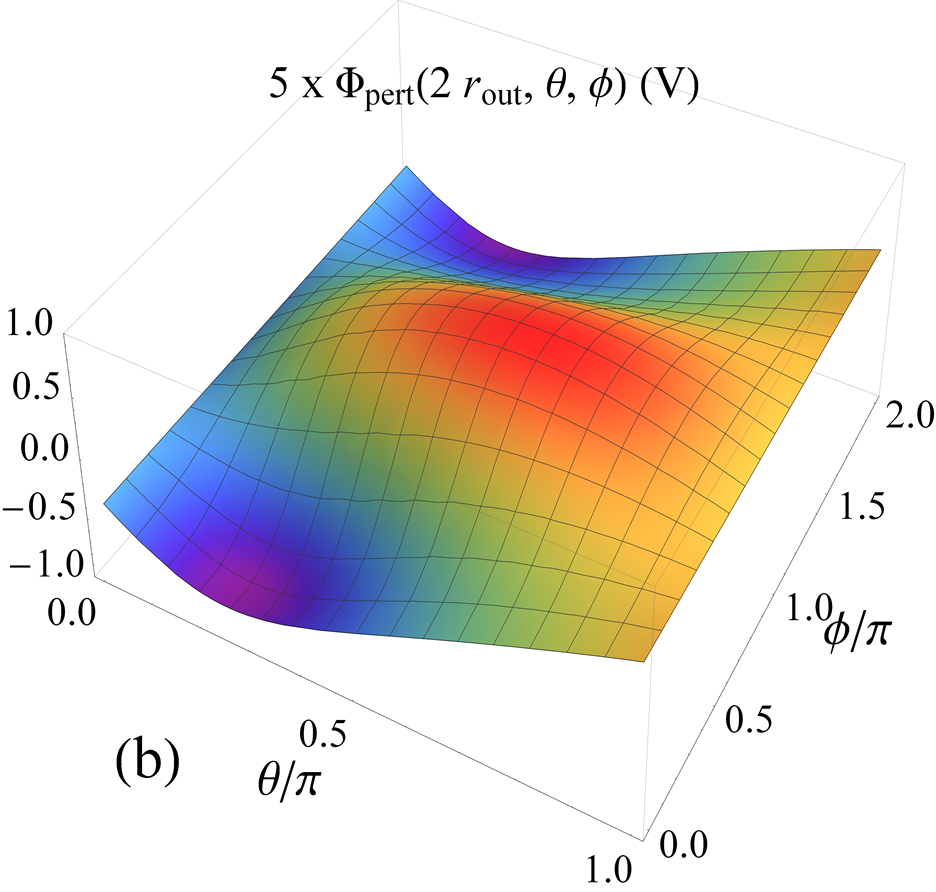} \\
\includegraphics[width=0.3\linewidth]{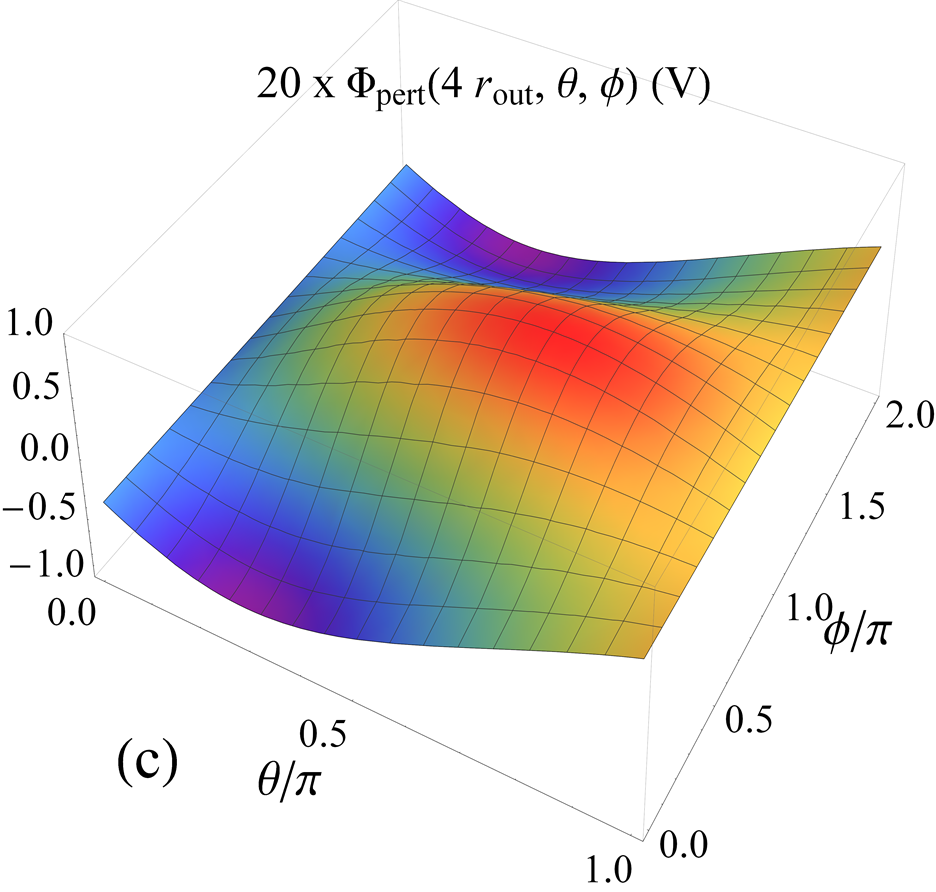}
\includegraphics[width=0.3\linewidth]{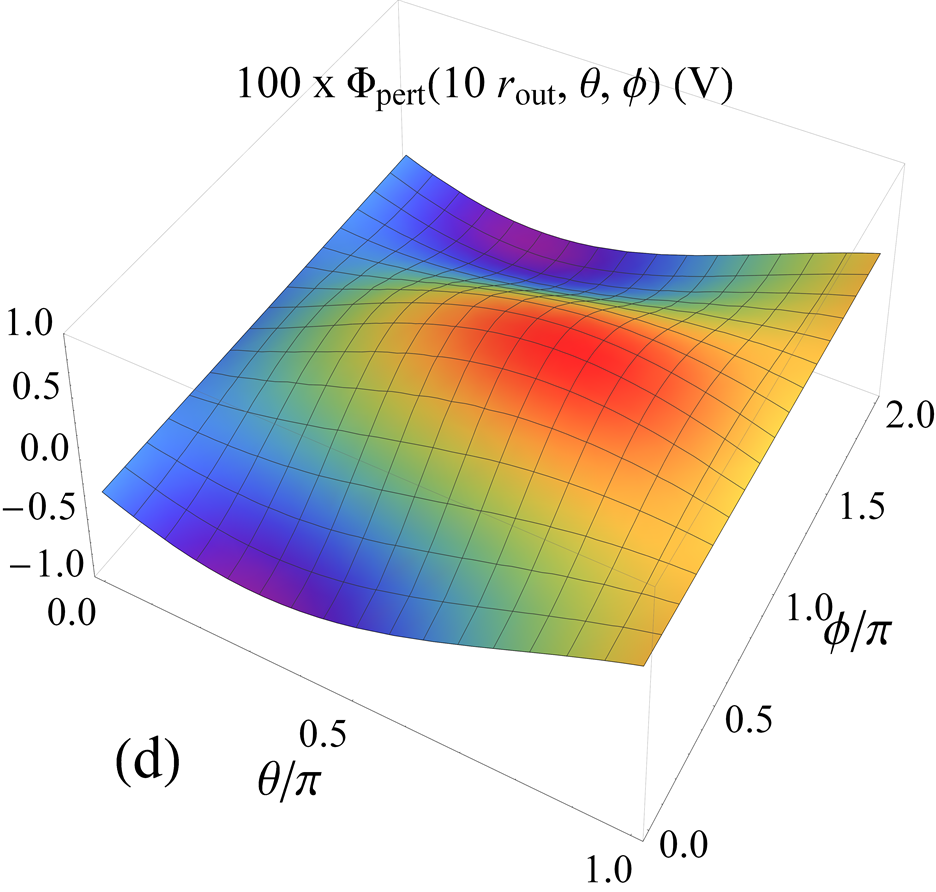}
\caption{Same as Fig. \ref{Fig3}, except that $\alphas = 2\pi/3$, $\betas = 3\pi/4$, and $\gammas = 5\pi/9$.}
\label{Fig4}
\end{figure}

Next, we considered the effect of $\=S \neq \=I$ (i.e., the shape principal axes are rotated with respect to the constitutive principal axes). We repeated the calculations of the previous case when $\alphas = 2\pi/3$, $\betas = 3\pi/4$, and $\gammas = 5\pi/9$; the results are depicted in Fig.~\ref{Fig4}. By comparing Figs.~\ref{Fig3} with \ref{Fig4}, it can be seen that the fact $\=S \neq \=I$ alters the rate of increase/ decrease of $\Phi_\text{pert}(r, \theta,\phi)$ in any specific direction. Furthermore, the locations of the extremums of $\Phi_\text{pert}(r, \theta,\phi)$
in the $\theta\phi$-plane are affected by the difference $\=S-\=I$
when $\rtilde <2$; see Figs. \ref{Fig3}(a) and \ref{Fig4}(a), for instance. The impact of $\=S - \=I$ diminishes as $\rtilde -2$ increases,
as can be seen on comparing Figs. \ref{Fig3}(c) and \ref{Fig4}(c), for instance.

%%%%%%%%%%%%%%%%  Figure 4 begins %%%%%%%%%%%%%%%%%%%%
%%%%%%%%%%%%%%%%  Figure 4 ends %%%%%%%%%%%%%%%%%%%%

\subsubsection{Point-dipole source}\label{pds}

\begin{figure}[H]
 \centering
\includegraphics[width=0.3\linewidth]{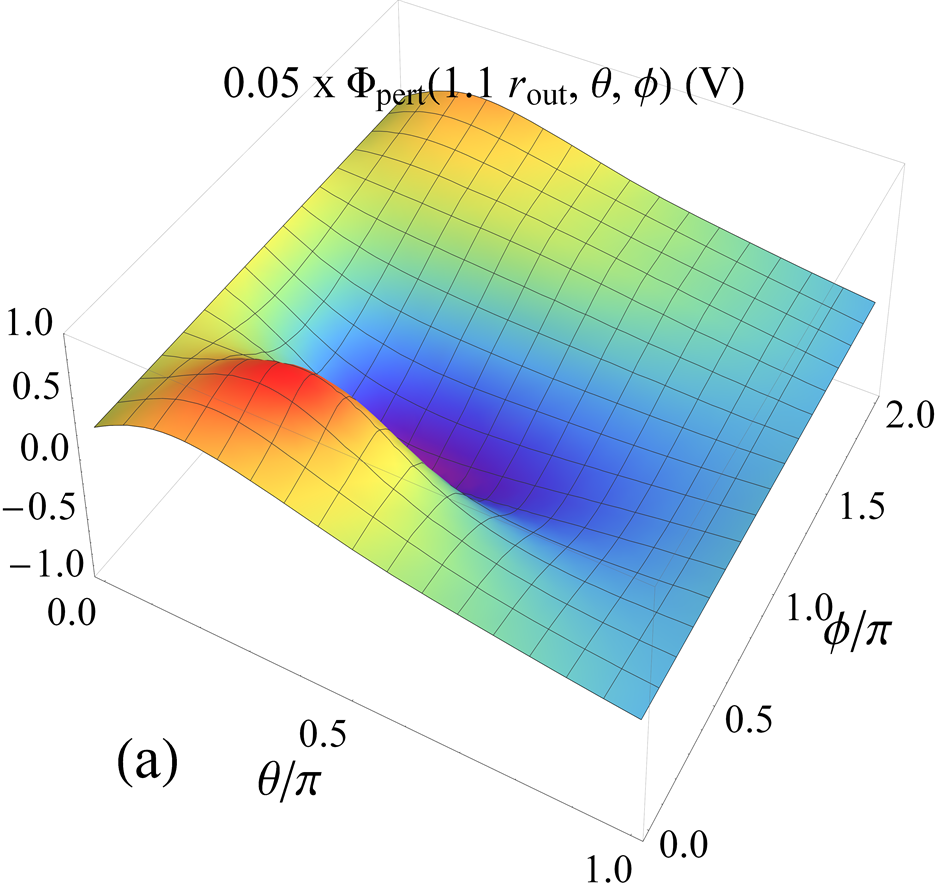}
\includegraphics[width=0.3\linewidth]{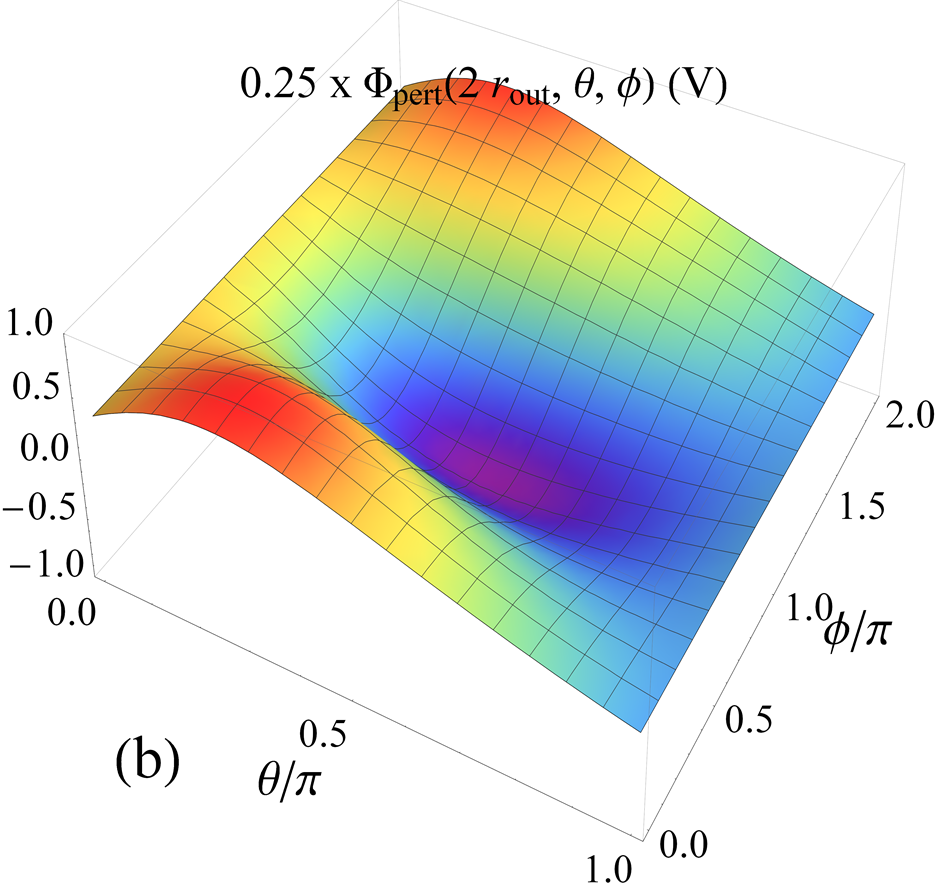} \\
\includegraphics[width=0.3\linewidth]{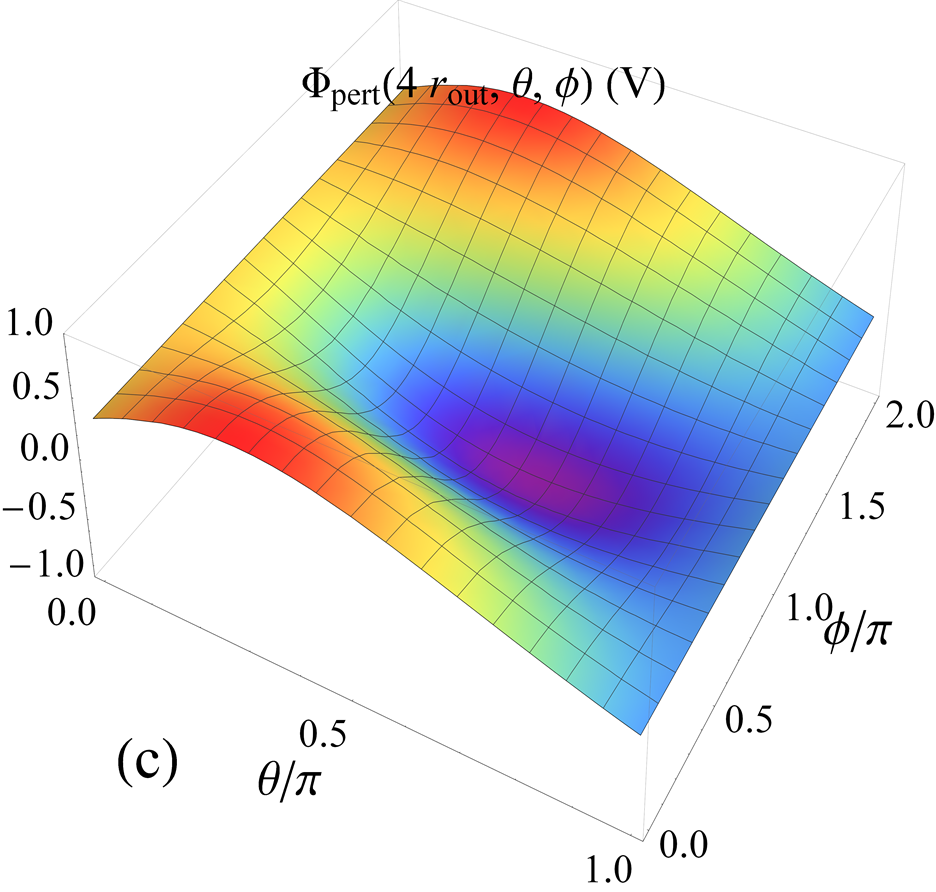}
\includegraphics[width=0.3\linewidth]{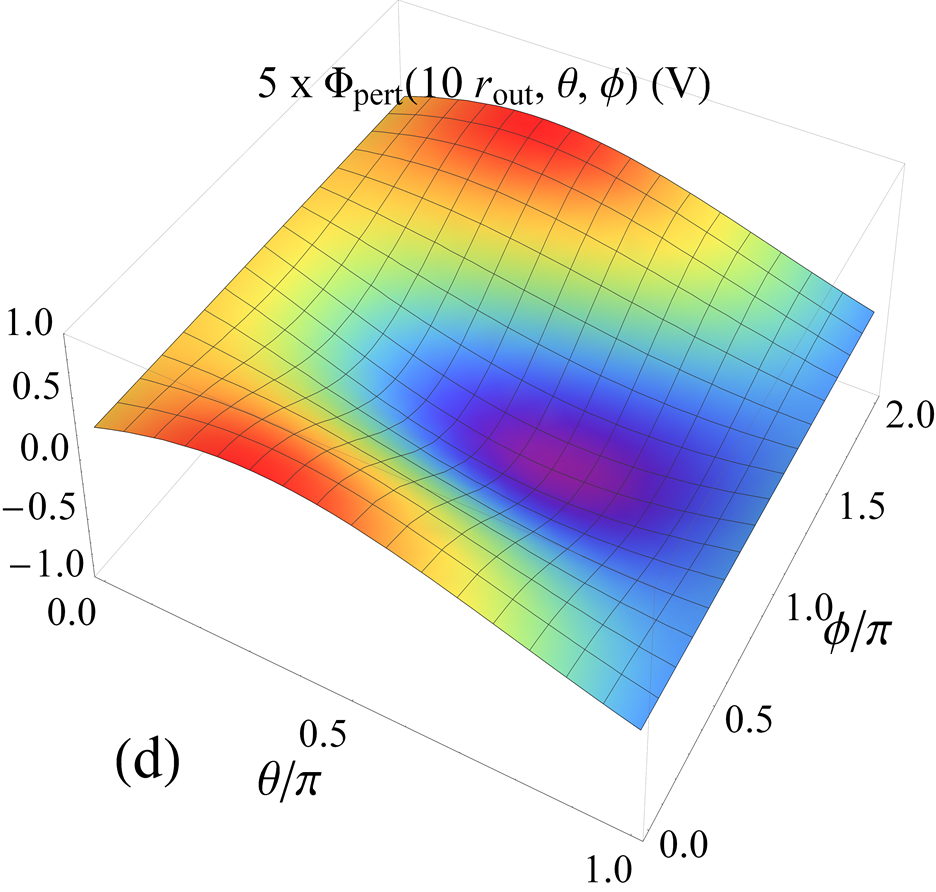}
\caption{Same as Fig. \ref{Fig3}, except for a point-dipole source with $p = 10^{-10}$~C~m. Also, $\thetap = \pi/4$ and $\phip = \pi/3$.}
\label{Fig5}
\end{figure}

Figures~\ref{Fig5}(a--d) present the angular profiles of  $\Phi_\text{pert}(\rtilde\,\rout, \theta,\phi)$ for $\rtilde \in \{ 1.1, 2, 4, 10\}$ for a point-dipole source with $p = 10^{-10}$~C~m, $\theta_p = \pi/4$, and $\phi_p = \pi/3$, when $\=S=\=I$. Just as with the point-charge source in Figs.~\ref{Fig3}(a--d),
$\Phi_\text{pert}(\rtilde\,\rout, \theta,\phi)$ decreases with increase of $\rtilde$; however, the rate of decrease for the point-dipole source  is smaller compared to that for the point-charge source. The effect of the type of source is visually evident on comparing Figs.~\ref{Fig3} and ~\ref{Fig5} for the locations of the extremums of $\Phi_\text{pert}(r, \theta,\phi)$ in the $\theta\phi$-plane.

\begin{figure}[H]
 \centering
\includegraphics[width=0.3\linewidth]{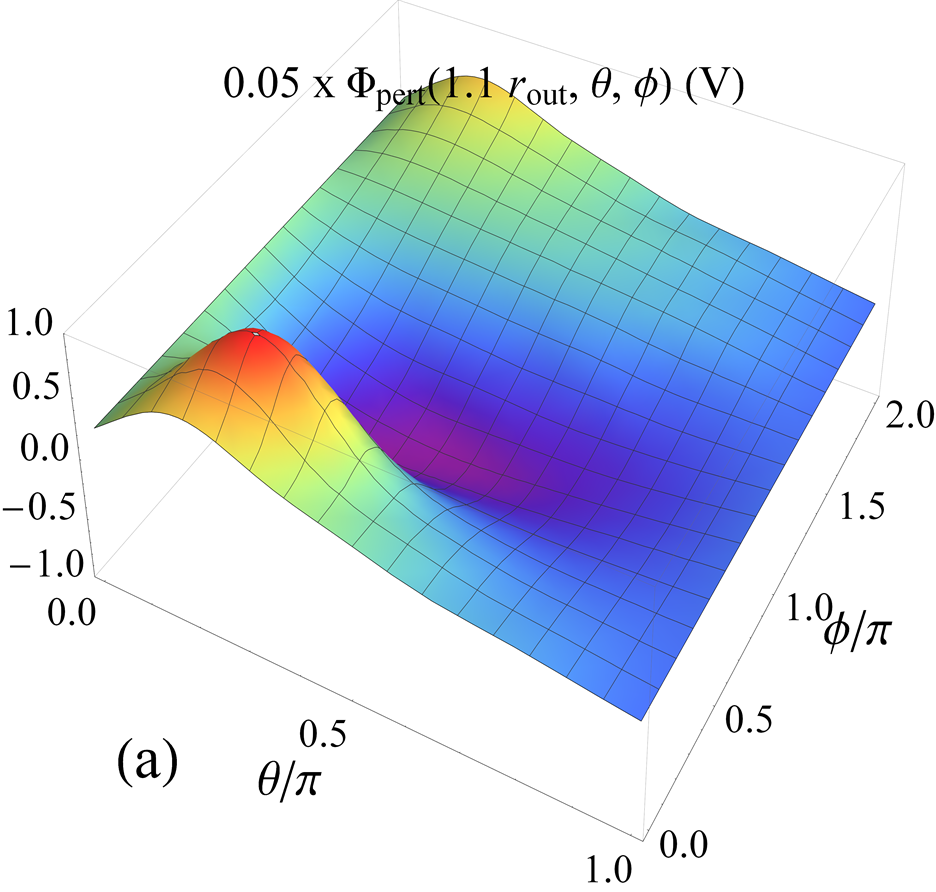}
\includegraphics[width=0.3\linewidth]{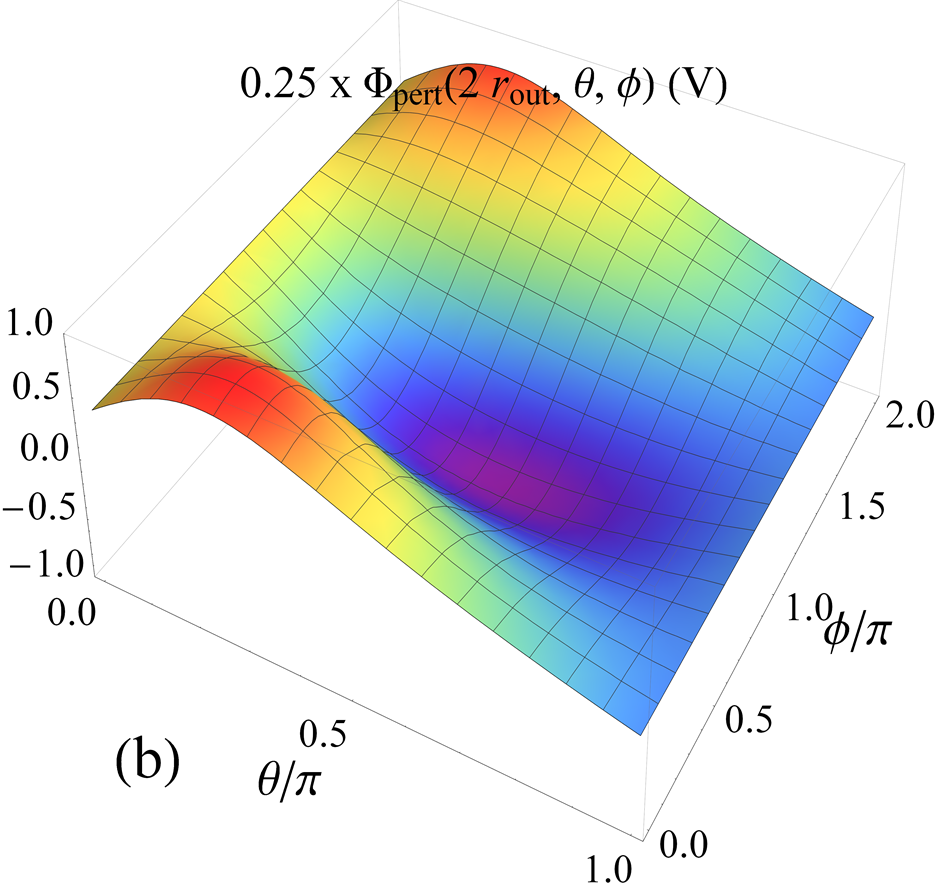} \\
\includegraphics[width=0.3\linewidth]{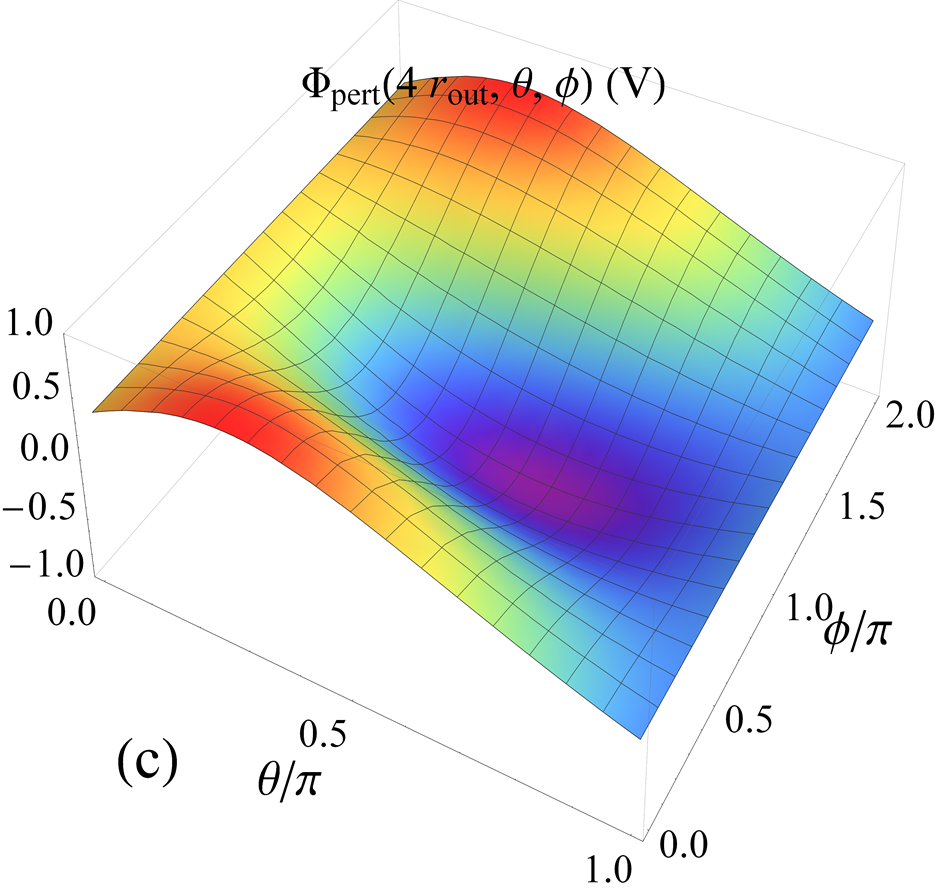}
\includegraphics[width=0.3\linewidth]{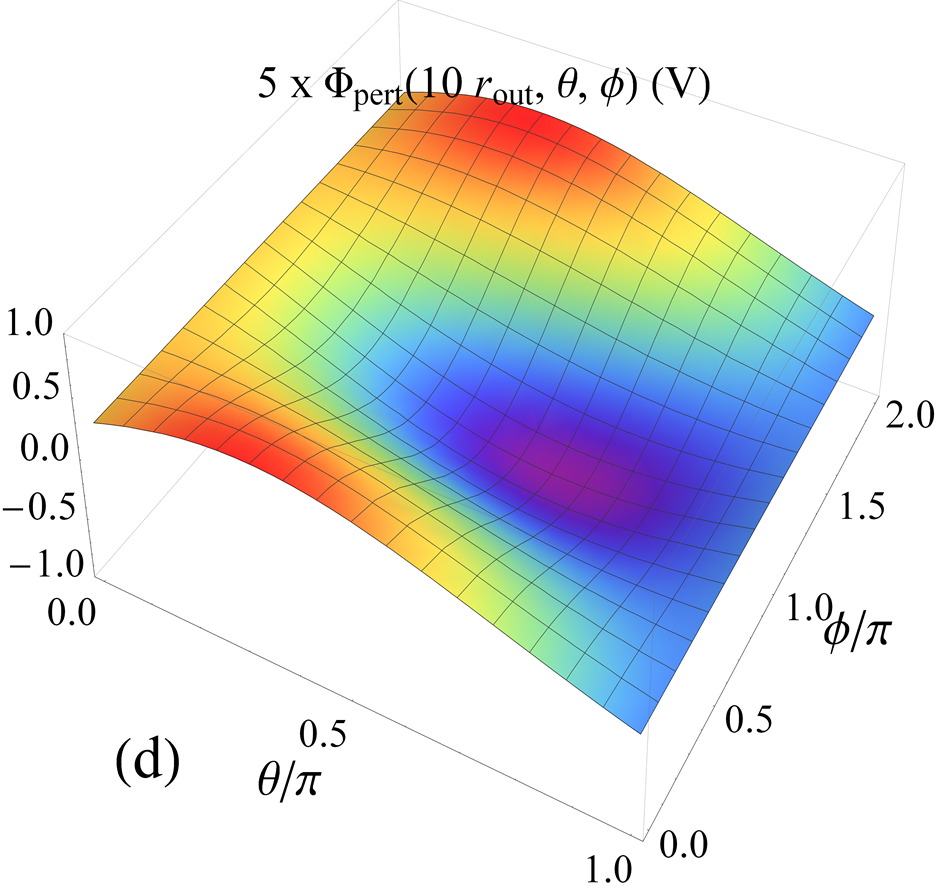}
\caption{Same as Fig. \ref{Fig5}, except when $\alphas = 2\pi/3$, $\betas = 3\pi/4$, and $\gammas = 5\pi/9$.}
\label{Fig6}
\end{figure}

Figures~\ref{Fig6}(a--d) present the same angular profiles as
Figs.~\ref{Fig5}(a--d), except that $\alphas = 2\pi/3$, $\betas = 3\pi/4$, and $\gammas = 5\pi/9$ (i.e., $\=S \neq \=I$). After comparing Figs.~\ref{Fig5} and \ref{Fig6}, we concluded that (similarly to Fig.~\ref{Fig4}) $\=S \neq \=I$ alters the increase/decrease rate of $\Phi_\text{pert}(r, \theta,\phi)$. As in Sec.~\ref{pcs}, the locations of the extremums of $\Phi_\text{pert}(r, \theta,\phi)$
in the $\theta\phi$-plane are mainly affected by the difference $\=S-\=I$
when $\rtilde <2$.

%%%%%%%%%%%%%%%%  Figure 5 begins %%%%%%%%%%%%%%%%%%%%
%%%%%%%%%%%%%%%%  Figure 5 ends %%%%%%%%%%%%%%%%%%%%

%%%%%%%%%%%%%%%%  Figure 6 begins %%%%%%%%%%%%%%%%%%%%
%%%%%%%%%%%%%%%%  Figure 6 ends %%%%%%%%%%%%%%%%%%%%

\section{Concluding Remarks}\label{conc}

We have formulated the EBCM  for the
perturbation of a source electric potential by a
 3D object  composed of a homogeneous anisotropic dielectric medium whose relative permittivity
dyadic is positive definite. The electrostatic counterpart of the Ewald--Oseen extinction theorem
was derived for that purpose, and
the electric potential inside the object was represented using a basis obtained by implementing
an affine bijective transformation of space to the Gauss equation for the electric field.  This formulation is
different from the Farafonov   formulation \cite{Farafonov2014,
FU2015,FIUP2016,Majic2017}, which is applicable only when the object is composed of a homogeneous
isotropic medium.

Even though the object is nonspherical, all potentials were represented in terms of the eigenfunctions of
the Laplace equation in the spherical coordinate system. Numerical results have shown that our EBCM
formulation fails to yield convergent results when the shape of the perturbing object deviates too much
from spherical (i.e., when either $\vert\mu-1\vert$ is substantial and/or $\vert\nu-1\vert$ is substantial). The
same problem bedevils the analogous formulation of the EBCM even for time-harmonic  problems \cite{Isk1,Isk2}.
As the Laplace equation is either separable or R-separable in several coordinate
systems \cite{MoonSpencer}, we plan to use  bases in future work that will conform better to the shape of the
object.

\vspace{5mm}
\noindent{\bf Acknowledgement.} AL is grateful to the Charles Godfrey Binder Endowment   for ongoing support
of his research activities.


\vspace{5mm}
\begin{thebibliography}{99}

\bibitem{Ewald1916}
P. P. Ewald, \quote{Zur Begr\"undung der Kristalloptik,}
\textit{Ann. Phys.}  \textbf{49},  117--143 (1916).

\bibitem{Oseen1915}
C. W. Oseen,   \quote{\"Uber die Wechselwirkung zwischen zwei elektrischen Dipolen und \"uber die Drehung der
Polarisationsebene in Kristallen und Fl\"ussigkeiten,}
\textit{Ann. Phys.}  \textbf{48},    1--56 (1915).

\bibitem{Lakh2018}
A. Lakhtakia, \quote{The Ewald--Oseen extinction theorem and the
extended boundary condition method,} in:  A. Lakhtakia and C. M. Furse (Eds.),
\textit{The World of Applied
Electromagnetics},  Springer, Cham, Switzerland,  2018;
 pp. 481--513.

\bibitem{Waterman1965}
P. C. Waterman,
\quote{Matrix formulation of electromagnetic scattering,}
\textit{Proc. IEEE} \textbf{53},  805--812 (1965).

\bibitem{Waterman1969EM}
P. C. Waterman,   \quote{Scattering by dielectric obstacles,}
\textit{Alta Freq. (Speciale)} \textbf{38},  348--352 (1969).

\bibitem{FarLakG4}
M. Faryad and A. Lakhtakia,
\textit{Infinite-Space Dyadic Green Functions in Electromagnetism},
Institute of Physics, Bristol, United Kingdom, 2018; Chap.~4.



\bibitem{Lin2004}
Z. Lin and S. T. Chui,
\quote{Electromagnetic scattering by optically anisotropic magnetic particle,}
\textit{Phys. Rev. E} \textbf{69},  056614 (2004).

\bibitem{Schmidt}
V. Schmidt and T. Wriedt,  \quote{The T-matrix for particle with arbitrary permittivity tensor
and parallelization of the computational code,}
\textit{J. Quant. Spectrosc.  Radiat.
Transf.} \textbf{113},  1712--1718 (2012).

\bibitem{Zouros}
G. P. Zouros, G. D. Kolezas, N. Stefanou, and T. Wriedt,
\quote{EBCM for electromagnetic modeling of gyrotropic BoRs,}
\textit{IEEE Trans. Antennas   Propagat.} \textbf{69}, (2021) doi: 10.1109/TAP.2021.3069589.

\bibitem{Doicu2019}
A. Doicu, T. Wriedt, and N. Khebbache,
\quote{An overview of the methods for deriving recurrence relations for T -matrix calculation,}
\textit{J. Quant. Spectrosc.  Radiat.
Transf.} \textbf{224},  1289--302 (2019).

\bibitem{Alkhoori3}
H. M. Alkhoori, A. Lakhtakia, J. K. Breakall, and
C. F. Bohren,
\quote{Plane-wave scattering by an ellipsoid composed
of an orthorhombic dielectric-magnetic
material with arbitrarily oriented
constitutive principal axes,} \textit{J. Opt. Soc. Am. B}
\textbf{36}, F60--F71 (2019).

\bibitem{Kahnert2020}
M. Kahnert and T. Rother,
\quote{Convergence of the iterative T-matrix method,}
\textit{Opt. Express} \textbf{28}, 28269--28282 (2020).




\bibitem{Jackson}
J. D. Jackson, \textit{Classical Electrodynamics}, 3rd ed.,
Wiley, Hoboken, NJ, USA,  1999; Sec.~1.8.

\bibitem{Colton}
D. Colton and R. Kress,
\textit{Integral Equation Methods in Scattering Theory}, SIAM, Philadelphia, PA, USA, 2013;
Theorem 3.1.

\bibitem{Farafonov2014}
V. G. Farafonov,
\quote{The Rayleigh hypothesis and the region of applicability
of the extended boundary condition method
in electrostatic problems for nonspherical particles,}
\textit{Opt. Spectrosc.} \textbf{117},  923--935 (2014).

\bibitem{Morse}
P. M. Morse and H. Feshbach,
\textit{Methods of Theoretical Physics, Vol. II}, McGraw--Hill,
New York, NY, USA,  1953; pp.~1264--1274.

\bibitem{FU2015}
V. G. Farafonov and V.~I. Ustimov,
\quote{Analysis
of the extended boundary condition method:
an electrostatic problem for Chebyshev particles,}
\textit{Opt. Spectrosc.} \textbf{118},  445--459 (2015).



\bibitem{FIUP2016}
V. G. Farafonov, V. Il'in,   V.~I. Ustimov, and M. Prokopjeva,
\quote{On the analysis of Waterman's approach
in the electrostatic case,}
\textit{J. Quant. Spectrosc. Radiat. Transf.} \textbf{178},  176--191 (2016).

\bibitem{Majic2017}
M. R. A. Maji\'{c}, F. Gray, B. Augui\'e, and E. C. Le~Ru,
\quote{Electrostatic limit of the T-matrix for electromagnetic scattering: Exact results for spheroidal particles,}
\textit{J. Quant. Spectrosc. Radiat. Transf.} \textbf{200},  50--58 (2017).

\bibitem{LTA-1}
A. Lakhtakia, N. L. Tsitsas, and H. M. Alkhoori, 
\quote{Theory of perturbation of electrostatic field by an  anisotropic dielectric sphere,}
arXiv:2108.06528 (2021).

\bibitem{Charnow}
A. Charnow and E. Charnow,
\quote{Fields for which the principal axis theorem is valid,}
\textit{Math. Mag.} \textbf{59},   222--225 (1986).

\bibitem{Strang}
G. Strang, \textit{Introduction to Linear Algebra}, 5th ed.,
Wellesley--Cambridge, Wellesley, MA, USA,  2016); p.~339.


\bibitem{Auld}
B. A. Auld, \textit{Acoustic Fields and Waves in Solids, Vol. I}, 2nd ed.,
Krieger, Malabar, FL, USA, 1990; p.~387.



\bibitem{MAEH}
T. G. Mackay and A. Lakhtakia, \textit{Modern Analytical Electromagnetic Homogenization
with Mathematica$^{\small \textregistered}$}, 2nd ed.,
Institute of Physics, Bristol, United Kingdom, 2020;
Sec.~6.5.

\bibitem{Smythe}
W. R. Smythe, \textit{Static and Dynamic Electricity}, 2nd ed.,
McGraw--Hill, New York, NY, USA, 1950; Sec.~3.09.



\bibitem{Tsitsas}
N.~L.~Tsitsas and P.~A.~Martin,
\quote{Finding a source inside a sphere,}
 \textit{Inverse Problems} \textbf{28},   015003 (2012).

 \bibitem{Redzic}
D. V. Red\v{z}i\'{c},
\quote{An electrostatic problem: A point charge outside a prolate dielectric spheroid,}
\textit{Am. J. Phys.} \textbf{62}, 1118--1121 (1994).

\bibitem{Isk1}
M.~F. Iskander, A. Lakhtakia, and C.~H. Durney,
\quote{A new procedure for improving the solution stability and
extending the frequency range of the EBCM,} \textit{IEEE Trans. Antennas Propagat.}
\textbf{31}, 317--324 (1983).

\bibitem{Isk2}
A. Lakhtakia, V.~K. Varadan, and V.~V. Varadan,
\quote{Iterative extended boundary condition method for scattering by
objects of high aspect ratios,} \textit{J. Acoust. Soc. Am.} \textbf{76}, 906--912 (1984).

\bibitem{MoonSpencer}
P. Moon and D. E. Spencer,
\quote{Separability conditions for the Laplace and Helmholtz equations,}
\textit{J. Franklin Inst.} \textbf{253}, 585--600 (1952).



\end{thebibliography}
\end{document}